# Deuterium chemistry and D/H ratios in Class 0/I proto-brown dwarfs

Riaz, B.[1⋆] & Thi, W.-F.[2]
[1] Universitäts-Sternwarte München, Ludwig Maximilians Universität, Scheinerstraße 1, 81679 München, Germany
[2] Max-Planck-Institut für Extraterrestrische Physik, Giessenbachstrasse 1, 85748 Garching, Germany

## ABSTRACT

We have conducted the first extensive observational survey of several deuterated species in 16 Class 0/I proto-brown dwarfs (proto-BDs) and 4 Class Flat/Class II brown dwarfs. Observations were obtained with the IRAM 30m telescope in the $DCO^+$ (3-2), DCN (3-2), DNC (3-2), and $N_2D^+$ (3-2) lines. The $DCO^+/H^{13}CO^+$, $DCN/H^{13}CN$, and $DNC/HN^{13}C$ ratios are comparatively higher and show a narrower range than the $DCO^+/HCO^+$, DCN/HCN, and DNC/HNC ratios, respectively. The mean D/H ratios for the proto-BDs derived from these molecules are in the range of ~0.02-3. Both low-temperature gas-phase ion-molecule deuteron transfer and grain surface reactions are required to explain the enhanced deuterium fractionation. The very dense and cold ($n_{H_2}$ $\geq 10^6$ cm$^{-3}$, T $\leq$10 K) interior of the proto-BDs provide the suitable conditions for efficient deuterium fractionation in these cores. There is no correlation between the D/H ratios and the CO depletion factor, with the exception of the DCN/HCN ratios that show a strong anti-correlation possibly due to the difference in the peak emitting regions of the DCN and HCN molecules. Over a wide range in the bolometric luminosities spanning ~0.002–40 $L_\odot$, we find a trend of higher $DCO^+/HCO^+$ (r = -0.7) and DCN/HCN (r = -0.6) ratios, nearly constant DNC/HNC (r = -0.4) and $DNC/HN^{13}C$ (r = -0.1) ratios, lower $N_2D^+/N_2H^+$ ratios (r = 0.6) in the proto-BDs compared to protostars. Only one Class II brown dwarf shows emission in the $DCO^+$ (3-2) line. No correlation is seen between the D/H ratios and the evolutionary stage.

**Key words:**
(stars:) brown dwarfs – stars: formation – stars: evolution – astrochemistry – ISM: abundances – ISM: molecules

## 1 INTRODUCTION

The ratio of deuterium to hydrogen (the D/H ratio), derived from the relative abundances of deuterated and non-deuterated isotopologues of molecules, is sensitive to the conditions in the physical environment, atmospheric enrichment, and chemical reprocessing, and thus can be an important tool to trace the conditions under which stars and planets form and evolve over time. The primordial elemental D/H ratio in the universe, set by the Big Bang nucleosynthesis calculations, is (2.8±0.2)×10$^{-5}$ (e.g., Pettini et al. 2008). The interstellar medium (ISM) of the Milky Way is somewhat depleted in deuterium relative to this value. Recent observations of the $HD/H_2$ species indicate that the mixing ratio of deuterium in the ISM shows a large scatter among different lines of sight, with a mean molecular D/H ratio of (2.0±0.1)×10$^{-5}$ (e.g., Sonneborn et al. 2000; Wood et al. 2004; Prodanović et al. 2010).

The enhancement of the molecular D/H ratio compared to the primordial or ISM elemental value, known as deuterium fractionation, is attributed to lower zero-point energies of the deuterated forms of the molecular ions that initiate the formation pathways of many ISM molecules (e.g., Tielens 1983; Dalgarno & Lepp 1984). While physical properties, most notably, low temperatures and high densities, can have a significant effect on deuterium fractionation, other factors such as the depletion factor of molecules, ionization degree, and metallicity can also play an important role. The sensitivity of the deuterium fractionation processes to these factors in the physical environment makes deuter-ated molecules a unique tracer to investigate the physics and chemistry of dense cores in dark clouds.

⋆ E-mail: briaz@usm.lmu.de

A variety of singly and doubly deuterated species have been detected in cold molecular clouds, pre-stellar cores, and low-mass protostars, including $DCO^+$, DCN, DNC, $H_2D^+$, $HD_2^+$, $N_2D^+$, $NHD_2$, $D_2H^+$ (e.g., Hirota et al. 2001; 2003; Loinard et al. 2000; 2002; Parise et al. 2002; 2006; 2011; Thi et al. 2004; Turner 2001; Jøergensen et al. 2004; Bacmann et al. 2003; Roueff & Gerin 2003; Bergin & Tafalla 2007; van der Tak et al. 2009; Taquet et al. 2014; Persson et al. 2018; Huang et al. 2017). The molecular D/H ratios derived from these species are 1%-10%, several orders of magnitude higher than the ISM D/H ratio. Both low-temperature gas-phase and surface grain chemistry can be attributed to the high levels of deuteration measured in cold molecular clouds, pre-stellar cores, and low-mass protostars (e.g., Roberts & Millar 2000ab; Loinard et al. 2002; Tielens 1983; Aikawa et al. 2003; 2012; Albertsson et al. 2013). Material enriched in deuterated species was then injected in the gas phase during the protostellar phase. Thermal evaporation due to heating by central protostar offers a natural pathway for re-injection in the gas phase. Non-thermal desorption mechanisms, such as, UV or cosmic ray radiation, or sputtering or partial destruction of the grains by fast shocks associated with high-velocity jets driven by protostars are known to be efficient in injecting dust grain material in the gas phase (e.g., Aikawa et al. 2012; 2018; Willacy et al. 2007; Garrod et al. 2007).

Deuterium has a notable importance in differentiating a brown dwarf from a giant planet. Substellar objects with masses above the limiting mass for thermonuclear fusion of deuterium are considered to be brown dwarfs, while objects with masses below this limit (and that orbit stars) are termed as planets. Brown dwarf evolutionary models place the limiting mass to distinguish between a brown dwarf and a giant planet to be 13±0.8 $M_{Jup}$ (e.g., Spiegel et al. 2011). This limiting mass is dependent on the initial D/H ratio, in addition to the initial He fraction and metallicity. In this context, it is important to determine the initial abundance of the various deuterated molecules and the physical conditions under which deuterium fractionation can take place in early-stage brown dwarfs. This can provide constraints on the initial physical and chemical conditions in the brown dwarf formation and evolutionary models. A study of deuterium fractionation in proto-BDs can also provide an insight into the freeze-out and deuteration timescales and how they compare with the infall timescale and the lifetime of the embedded stage in these objects.

We have conducted the first extensive observational survey of several deuterated species in Class 0/I brown dwarfs, also termed as proto-brown dwarfs (hereafter; proto-BDs), with an aim to measure the D/H ratio for brown dwarfs during their earliest evolutionary stages, and investigate how the D/H ratios compare with other cold dense objects and planets and comets in the solar system. Proto-BDs are high-density ($n_{H_2}$ $\geq 10^6$ $cm^{-3}$) and cold ($T_{kin}$ $\leq$10 K) objects (Machida et al. 2009). Since the deuterium enrichment of molecules is sensitive to their formation environment, this makes the interiors of proto-BDs ideal to study deuterium fractionation. The sample of proto-BDs studied in this work and the IRAM 30m observations are described in Section 2. Results on the various deuterated species detected in the proto-BDs and the formation pathways of the species are presented in Sect. 4. Section 5 presents a comparison of the D/H ratios for the proto-BDs with low-mass protostars (Sect. 5.1), other astronomical objects (Sect. 5.2), and the predictions from chemical models (Sect. 5.3). A discussion on the deuteration and evolutionary timescales is presented in Sect. 5.5.

## 2 TARGETS, OBSERVATIONS AND DATA REDUCTION

### 2.1 Sample

Our sample consists of 20 targets, 7 of which are located in Serpens, 5 in Perseus, 4 in Ophiuchus, and 1 in the Taurus region. Table 1 lists the properties for the targets in the sample. We applied the same target selection and identification criteria as described in Riaz et al. (2015). The total (dust+gas) mass derived from the JCMT SCUBA-2 sub-millimeter (sub-mm) 850μm flux density is in the range of ~0.004–0.065 $M_\odot$ for the targets. The bolometric luminosity, $L_{bol}$, for the targets is in the range of ~0.004–0.1 $L_\odot$. This is the extinction-corrected $L_{bol}$ measured $L_{bol}$ from integrating the observed infrared to sub-mm SEDs. The internal luminosity, $L_{int}$, is estimated to be ~70%-80% of $L_{bol}$. We assumed a distance of 436±9 pc to the Serpens region, 144±6 pc to Ophiuchus, 294±17 pc to Perseus, and 145±10 pc to Taurus (Ortiz-Leon et al. 2017; Dzib et al. 2010; Mamajek 2008; Schlafly et al. 2014; Zucker et al. 2019).

As discussed in detail in Riaz et al. (2016), both $L_{bol}$ and $L_{int}$ for these targets are below the luminosity threshold typically considered between very low-mass stars and brown dwarfs (<0.1 $L_\odot$) in brown dwarf evolutionary models and the more recent accretion models (e.g., Baraffe et al. 2017). This indicates the sub-stellar nature of the targets. We further note that the *present* $M^{d+g}_{total}$ is sub-stellar and about 30%-50% of the mass is expected to be expelled by jets/outflows during the main accretion phase (e.g., Machida et al. 2009). Therefore, the *final* mass for these targets is also expected to stay below the sub-stellar limit.

A Class 0/I object would still be in the main accretion phase. While accretion rate measurements are not available for all of these targets, from Riaz & Bally (2021), the ratio of the outflow to accretion rate shows a wide range of ~0.01-0.8 for proto-BDs. The main point to note is that the central object mass for these very low $L_{bol}$ cases is typically ~10 $M_{Jup}$ or even less (Riaz & Machida 2021). Also, the total circumstellar mass is very low. Thus, even if all of the circumstellar mass is accreted onto the central object, the total mass will stay below the sub-stellar limit.

We can do a simple calculation to estimate the accretion from a giant molecular cloud such as Ophiuchus or Serpens onto a proto-BD core, assuming that the object could still be accreting from the wider environment via e.g., filament accretion. If we assume a slow, steady flow of gas from a low-density cloud to the proto-BD core, then for a velocity of ~0.1 km/s, a core mass of 0.01 $M_\odot$, T=10 K, $n(H_2) = 10^4$ $cm^{-3}$, then the accretion rate is estimated to be ~2× $10^{-9}$ $M_\odot$/yr. This implies that a 0.01 $M_\odot$ core will gain ~2 $M_{Jup}$ in 1 Myr. This is an insignificant gain in the final mass of the core due to accretion from the surrounding cloud.

The 2-24 μm spectral slope for 16 targets is >0.3, con-sistent with a Class 0/I classification, while the remaining 4 have SED slopes between 0.3 and -1, thus classifying them as Class Flat/Class II objects. We have also determined the evolutionary stage of these YSOs using the Stage 0+I/II criteria based on the strength in the $HCO^+$ (3-2) line emission, and the Stage 0/I/I-T/II classification criteria based on the physical characteristics of the system. A YSO with an integrated intensity of >0.4 K km $s^{-1}$ in the $HCO^+$ (3-2) line is classified as a Stage 0+I object (e.g., Riaz et al. 2016). Based on the physical characteristics estimated from the radiative transfer modelling of the observed spectral energy distribution (SED), a Stage 0 object is expected to have a disk-to-envelope mass ratio of <<1, while for a Stage I YSO, this mass ratio would be in the range of 0.1–2 (e.g., Whitney et al. 2003). A Stage 0 object is also expected to have a total circumstellar (disk+envelope) mass to stellar mass ratio of ~1. The Stage thus determined has been compared to the Class 0/I/Flat classification based on the 2-24 μm slope of the observed SED. Stage I-T and Class Flat are considered to be intermediate between the Stage I and Stage II, or Class I and Class II evolutionary phases.

As listed in Table 1, we find a good match between the Stage of the system based on the physical characteristics and the strength in the molecular line emission, whereas the classification based on the SED slope does not relate well to this evolutionary stage. An edge-on inclination could result in a flatter spectral slope in the SED compared to genuine Stage 0/I systems (e.g., Riaz et al. 2016; 2021). Based on these criteria, our sample consists of 16 Class 0/I proto-BDs, half of which are in Stage 0 and half in Stage I (Table 1). The remaining 4 objects are more evolved Stage I-T/II brown dwarfs.

### 2.2 Molecular Line Observations and Data Reduction

Observations were obtained in various observing runs at the IRAM 30m telescope between the years 2017 and 2020. We used the EMIR heterodyne receiver and three tuning setups in the E230 band to observe the $DCO^+$ (3-2), DCN (3-2), DNC (3-2), $N_2D^+$ (3-2), and $N_2H^+$ (3-2) lines. An overview of the molecular lines observed within these frequency ranges are listed in Table 2. Due to poor weather conditions, the $N_2H^+$ (3-2) line observations could only be obtained for 8 targets. We used the FTS backend in the wide mode, with a spectral resolution of 200 kHz (~0.3 km $s^{-1}$ at 224 GHz). The observations were taken in the frequency switching mode with a frequency throw of approximately 7 MHz. The source integration times ranged from 3 to 4 hours per source per tuning reaching a typical RMS (on $T_A^*$ scale) of ~0.01-0.02 K. The telescope absolute RMS pointing accuracy is better than 3″ (Greve et al. 1996). All observations were taken under good weather conditions (0.08 <τ <0.12; PWV <2.5 mm). The absolute calibration accuracy for the EMIR receiver is around 10% (Carter et al. 2012). The telescope intensity scale was converted into the main beam brightness temperature ($T_{mb}$) using standard beam efficiency of ~59% at 230 GHz. The half power beam width of the telescope beam is 10″ at 230 GHz. The spectral reduction was conducted using the CLASS software (Hily-Blant et al. 2005) of the GILDAS facility[1]. The standard data reduction process consisted of averaging multiple observations for each transition line, extracting a subset around the line rest frequency, and finally a low-order polynomial baseline was subtracted for each spectrum.

## 3 DATA ANALYSIS

The observed $DCO^+$ (3-2), DCN (3-2), DNC (3-2), $N_2D^+$ (3-2), $N_2H^+$ (3-2) spectra are shown in Figs. A1-A5, respectively, in the Appendix A (Supplementary Material). Most spectra for the deuterated species show a single-peaked Gaussian profile with no self-absorption at the cloud systemic velocity. We have measured the parameters of the line center, line width, the peak and integrated intensities using a single- or double-peaked peaked Gaussian fit. The line parameters derived from the $DCO^+$ (3-2), DCN (3-2), DNC (3-2), $N_2D^+$ (3-2), $N_2H^+$ (3-2) spectra are listed in Tables A1-A5, respectively, in Appendix A. For the non-detections, a 2-σ upper limit was calculated by integrating over the velocity range of ±2 km $s^{-1}$ from the cloud systemic velocity of ~8 km $s^{-1}$ for Serpens and Perseus, and ~4 km $s^{-1}$ for Ophiuchus and Taurus (e.g., White et al. 2015; Burleigh et al. 2013). We estimate uncertainties of ~15%-20% for the integrated intensity, ~6%-8% on $V_{lsr}$, ~5%-10% on $T_{mb}$ and Δv.

As a check, we also obtained off-source spectra in the $DCO^+$ (3-2) line for the Serpens and Ophiuchus targets. These off-source spectra were taken at a step size of one beamsize (~10″) eastward from the target position. The off-source flux in the $DCO^+$ (3-2) line is <3-σ (1-σ rms ~ 0.02 K). For the cases with a weak peak $DCO^+$ line intensity of ~0.1-0.2 K (e.g., J163152, J182854), the off-source flux is ~0.02-0.03 K, while for objects like J162625, the off-source flux is ~0.05-0.06 K. The off-source flux is thus quite weak and about the same as the 2-σ rms. Since we do not have off-source spectra in all of the lines for all targets, we have not applied a background subtraction to all spectra. Future observations of such spectra at various off-source positions, as well as maps of the surroundings for all targets in various deuterium species can provide a more robust estimate on the emission that arises from the surrounding cloud.

We have used the non-LTE radiative transfer code RADEX (van der Tak et al. 2007) to estimate the column densities for all species detected towards the proto-BDs. For a fixed kinetic temperature, $T_{kin}$, $H_2$ number density, $n_{H_2}$, and line width in RADEX, the input column density is varied to match with the observed peak line intensity ($T_{mb}$). We set $T_{kin}$ to 10 K, and derived $H_2$ number and column density from the total mass $M_{total}^{d+g}$ and the outer radius of the envelope, $R_{env}$, as described in Riaz et al. (2018). The range in $M_{total}^{d+g}$ is ~20-60 $M_{Jup}$ and ~400-800 au for $R_{env}$. The output from RADEX is the radiation temperature $T_R$, which was converted to $T_{mb}$ by making corrections for the main beam efficiency and the beam dilution factor. The column densities and molecular abundances relative to $H_2$, [N(X)/N($H_2$)], for the deuterated and non-deuterated species are listed in Table 3 and Table 4, respectively.

[1] http://www.iram.fr/IRAMFR/GILDAS

**Table 1.** Sample

| Object | RA (J2000) | Dec (J2000) | $L_{bol}$ ($L_\odot$)[a] | Classification[b] | Region |
|---|---|---|---|---|---|
| SSTc2d J182854.9+001833 (J182854) | 18h28m54.90s | 00d18m32.68s | 0.05 | Stage 0+I, Stage I, Class 0/I | Serpens |
| SSTc2d J182844.8+005126 (J182844) | 18h28m44.78s | 00d51m25.79s | 0.04 | Stage 0+I, Stage 0, Class 0/I | Serpens |
| SSTc2d J183002.1+011359 (J183002) | 18h30m02.09s | 01d13m58.98s | 0.09 | Stage 0+I, Stage 0, Class 0/I | Serpens |
| SSTc2d J182959.4+011041 (J182959) | 18h29m59.38s | 01d10m41.08s | 0.008 | Stage 0+I, Stage I, Class 0/I | Serpens |
| SSTc2d J182953.0+003607 (J182953) | 18h29m53.05s | 00d36m06.72s | 0.12 | Stage 0+I, Stage 0, Class Flat | Serpens |
| SSTc2d J182856.6+003008 (J182856) | 18h28m56.64s | +00d30m08.30s | 0.004 | Stage 0+I, Stage 0, Class 0/I | Serpens |
| SSTc2d J182952.2+011559 (J182952) | 18h29m52.21s | +01d15m59.10s | 0.024 | Stage 0+I, Stage 0, Class 0/I | Serpens |
| SSTc2d J163143.8-245525 (J163143) | 16h31m43.75s | -24d55m24.61s | 0.09 | Stage 0+I, Stage I, Class Flat | Ophiuchus |
| SSTc2d J163136.8-240420 (J163136) | 16h31m36.77s | -24d04m19.77s | 0.09 | Stage 0+I, Stage I, Class Flat | Ophiuchus |
| SSTc2d J163152.06-245726.0 (J163152) | 16h31m52.06s | -24d57m26.0s | 0.009 | Stage 0+I, Stage 0, Class 0/I | Ophiuchus |
| SSTc2d J162625.62-242428.9 (J162625) | 16h26m25.62s | -24d24m28.9s | 0.04 | Stage 0+I, Stage 0, Class 0/I | Ophiuchus |
| SSTc2d J032838.78+311806.6 (J032838) | 03h28m38.78s | +31d18m06.6s | 0.017 | Stage 0+I, Stage I, Class 0/I | Perseus |
| SSTc2d J032848.77+311608.8 (J032848) | 03h28m48.77s | +31d16m08.8s | 0.013 | Stage 0+I, Stage I, Class 0/I | Perseus |
| SSTc2d J032851.26+311739.3 (J032851) | 03h28m51.26s | +31d17m39.3s | 0.011 | Stage 0+I, Stage 0, Class 0/I | Perseus |
| SSTc2d J032859.23+312032.5 (J032859) | 03h28m59.23s | +31d20m32.5s | 0.006 | Stage 0+I, Stage I, Class 0/I | Perseus |
| SSTc2d J032911.89+312127.0 (J032911) | 03h29m11.89s | +31d21m27.00s | 0.06 | Stage 0+I, Stage I, Class 0/I | Perseus |
| SSTtau 041858.1+281223 (J041858) | 04h18m58.0s | +28d12m23.0s | 0.04 | Stage II, Stage II, Class II | Taurus |
| SSTc2d J182940.2+001513 (J182940) | 18h29m40.20s | 00d15m13.11s | 0.074 | Stage II, Stage I-T, Class 0/I | Serpens |
| SSTc2d J182927.4+003850 (J182927) | 18h29m27.35s | 00d38m49.75s | 0.012 | Stage II, Stage I-T, Class 0/I | Serpens |
| SSTc2d J182952.1+003644 (J182952) | 18h29m52.06s | 00d36m43.63s | 0.016 | Stage II, Stage II, Class Flat | Serpens |

[a] Errors on $L_{bol}$ is estimated to be ~20%.
[b] The first, second, and third values are using the classification criteria based on the integrated intensity in the $HCO^+$ (3-2) line, the physical characteristics, and the SED slope, respectively.

**Table 2.** IRAM 30m Molecular line observations

| Molecule | Line | Frequency (GHz) | $E_j/k$ (K) | $A_{jk}$ ($s^{-1}$) | $n^{thin}_{crit}$(10 K) ($cm^{-3}$) |
|---|---|---|---|---|---|
| $DCO^+$ | 3-2 | 216.1126 | 20.7 | $7.7\times10^{-4}$ | $2.0\times10^6$ |
| DCN | 3-2 | 217.2386 | 20.8 | $4.6\times10^{-4}$ | $2.8\times10^7$ |
| DNC | 3-2 | 228.9105 | 21.9 | $5.5\times10^{-4}$ | $4.8\times10^6$ |
| $N_2D^+$ | 3-2 | 231.3218 | 22.2 | $7.1\times10^{-4}$ | $1.8\times10^6$ |
| $N_2H^+$ | 3-2 | 279.5117 | 26.8 | $12.6\times10^{-4}$ | $3.7\times10^6$ |

We estimate an uncertainty of ~23-32% on the column densities and abundances. The uncertainties have been determined from propagating the error on the integrated line intensity and the $H_2$ column densities. For the non-detections, we have derived the upper limits on the column density and abundance using the 2-$\sigma$ upper limit on the integrated intensity, assuming a line width of 1 km $s^{-1}$ and the same kinetic temperature. The output from RADEX also includes the excitation temperature $T_{ex}$ and the optical depth $\tau$. These are listed in Table A6 (Appendix A in the Supplementary Material).

## 4 RESULTS

Among the Stage 0/I proto-BDs, 15 show emission in $DCO^+$ (3-2), 7 in DCN (3-2), 9 in DNC (3-2), and 5 in $N_2D^+$ (3-2). No dependence of the D/H ratios is seen on the star-forming region (Table 5); however, the highest CO depletion factors are measured for the Perseus targets, indicating that these sources are embedded in a dense environment. Several factors can be responsible for the non-detections, such as, some of the proto-BDs may be more chemically evolved than the others, or the destruction of deuterated molecules via grain-surface reactions that may result in a non-detection, or enhanced chemical processing once molecules are released in the gas-phase, or the abundances could be much below the detection limit [$N(X)/N(H_2)<1\times10^{-12}$]. The non-detection could also be due to a lower initial D abundance, or primordial D abundance. The share distributed in the host cloud could be low, or a low fraction was accreted onto the central proto-BD core if formed via core accretion. Some of the proto-BDs may have formed in warmer environment than others.

In the following section, we first present the formation pathways for the deuterated species and then the results on each molecule for the proto-BDs.

### 4.1 Formation pathways of deuterated species

Chemical models predict the enrichment of HD, the main deuterium reservoir, via ion-molecule reactions with polyatomic ions, such as $H_3^+$ (and $H_2D^+$, $HD_2^+$ , $D_3^+$).

$$HD + H_3^+ \rightarrow H_2D^+ + H_2 + 232\ \mathrm{K} \quad (1)$$

The backward reactions between $H_2$ and $H_3^+$ isotopologues are endothermic by ~100–300 K, leading to the initial

Table 3: Column densities and molecular abundances for deuterated species

| Object | $n_{H_2}$ x$10^6$ ($cm^{-3}$) | $N_{H_2}$ x$10^{22}$ ($cm^{-2}$) | N($DCO^+$) x$10^{11}$ ($cm^{-2}$) | [$DCO^+$] x$10^{-11}$ | N(DCN) x$10^{11}$ ($cm^{-2}$) | [DCN] x$10^{-11}$ | N(DNC) x$10^{11}$ ($cm^{-2}$) | [DNC] x$10^{-11}$ | N($N_2D^+$) x$10^{11}$ ($cm^{-2}$) | [$N_2D^+$] x$10^{-11}$ |
|---|---|---|---|---|---|---|---|---|---|---|
| J182854 | 1.0±0.4 | 3.7±0.6 | 1.1 | 0.3 | <4 | <1 | <3 | <0.8 | <1 | <0.3 |
| J182844 | 0.7±0.3 | 3.6±0.5 | 15.0 | 4.2 | 7.0 | 2.0 | 60.0 | 16.0 | 11.0 | 3.0 |
| J183002 | 2.1±0.2 | 7.6±0.5 | 3.1 | 0.4 | 0.3 | 0.04 | 2.0 | 0.3 | <0.8 | <0.1 |
| J182959 | 0.8±0.5 | 4.5±0.5 | 30.0 | 6.6 | 4.0 | 0.8 | 22.0 | 4.8 | 3.0 | 0.6 |
| J163143 | 0.3±0.2 | 1.8±0.2 | 7.6 | 4.2 | <6 | <3 | <7 | <4 | <1 | <0.5 |
| J182953 | <0.5 | <2.0 | 1.7 | 0.8 | <6 | <3 | <6 | <3 | <1 | <0.5 |
| J163136 | 0.6±0.1 | 2.9±0.9 | <0.7 | <0.2 | <4 | <1 | <4 | <1 | <1 | <0.3 |
| J182856 | 1.4 ±0.2 | 3.4±0.4 | 10.3 | 3.4 | 5.5 | 1.6 | 1.6 | 0.5 | 1.2 | 0.3 |
| J182952 | 3.6 ±0.3 | 8.7±0.6 | 10.2 | 1.1 | 2.5 | 0.3 | 6.3 | 0.7 | 4.6 | 0.5 |
| J163152 | 1.5 ±0.2 | 1.0±0.2 | 0.8 | 0.8 | <1 | <1 | <1 | <1 | <0.4 | <0.4 |
| J162625 | 3.3 ±0.5 | 4.0±0.6 | 80.5 | 20.3 | 5.5 | 1.4 | – | – | 8.2 | 2.3 |
| J032838 | 8.6±1.2 | 21.3±3.0 | 7.5 | 0.3 | <0.3 | <0.01 | 1.1 | 0.05 | <0.8 | <0.04 |
| J032848 | 4.2±0.6 | 10.4±1.5 | 2.2 | 0.2 | <0.5 | <0.05 | 2.4 | 0.2 | 2.4 | 0.2 |
| J032851 | 3.0±0.4 | 7.3±2.2 | 1.5 | 0.2 | – | – | <0.5 | <0.07 | 0.5 | 0.07 |
| J032859 | 14.2±2 | 34.2±12.4 | 15.3 | 0.4 | 2.4 | 0.07 | 7.4 | 0.2 | – | – |
| J032911 | 5.2 ±1.4 | 12.5±0.8 | 0.9 | 0.07 | <0.5 | <0.04 | <1.5 | <0.1 | <0.4 | <0.03 |
| J041858 | 1.3 ±0.2 | 3.2±0.4 | 4.0 | 1.2 | <1 | <0.3 | 3.3 | 0.9 | <0.3 | <0.09 |
| J182940 | 0.05±0.04 | 0.4±0.07 | <0.6 | <1.5 | <40 | <100 | <50 | <100 | <6 | <15 |
| J182927 | <0.1 | <0.6 | <0.8 | <1.3 | <30 | <50 | <40 | <60 | <5 | <8 |
| J182952 | <0.9 | <3 | <0.7 | <0.2 | <3 | <1 | <3 | <1 | <0.8 | <0.3 |

The uncertainty is estimated to be ~23-32% for the column densities and molecular abundances. A '–' indicates that observations are not available.

Table 4: Column densities and molecular abundances for non-deuterated species

| Object | $N(H^{13}CO^+)$ x$10^{11}$ ($cm^{-2}$) | $[H^{13}CO^+]$ x$10^{-11}$ | $N(H^{13}CN)$ x$10^{11}$ ($cm^{-2}$) | $[H^{13}CN]$ x$10^{-11}$ | $N(HN^{13}C)$ x$10^{11}$ ($cm^{-2}$) | $[HN^{13}C]$ x$10^{-11}$ | $N(HCO^+)$ x$10^{11}$ ($cm^{-2}$) | $[HCO^+]$ x$10^{-11}$ | N(HCN) x$10^{11}$ ($cm^{-2}$) | [HCN] x$10^{-11}$ | N(HNC) x$10^{11}$ ($cm^{-2}$) | [HNC] x$10^{-11}$ | $N(N_2H^+)$ x$10^{11}$ ($cm^{-2}$) | $[N_2H^+]$ x$10^{-11}$ |
|---|---|---|---|---|---|---|---|---|---|---|---|---|---|---|
| J182854 | 1.4 | 0.4 | <8.0 | <2.0 | <8.0 | <2.0 | 22.0 | 6.0 | 15.0 | 2.0 | 65.0 | 17.0 | 50.5 | 13.2 |
| J182844 | 5.5 | 1.5 | <20.0 | <5.0 | <10.0 | <3.0 | 24.0 | 6.6 | 34.0 | 9.0 | 69.0 | 19.0 | 270.2 | 90.3 |
| J183002 | 4.0 | 0.5 | <5.0 | <0.6 | <5.0 | <0.6 | 345.0 | 45.0 | – | – | 85.0 | 12.0 | 300.2 | 40.0 |
| J182959 | 6.5 | 1.4 | <30.0 | <6.0 | 7.0 | 1.5 | 70.0 | 15.0 | <20.0 | <4.0 | 120.0 | 26.0 | 130.4 | 32.4 |
| J163143 | <0.1 | <0.05 | <40.0 | <20.0 | <30.0 | <20.0 | 250.0 | 138.0 | <20.0 | <11.0 | 60.0 | 33.0 | – | – |
| J182953 | 6.0 | 3.0 | <20.0 | <10.0 | <20.0 | <10.0 | 200.0 | 100.0 | <20.0 | <10.0 | 80.0 | 40.0 | – | – |
| J163136 | <0.05 | <0.02 | <20.0 | <7.0 | <20.0 | <7.0 | 15.0 | 5.0 | <20.0 | <7.0 | <20.0 | <7.0 | – | – |
| J182856 | 5.0 | 1.5 | <4.0 | <1.2 | <3.5 | <1.0 | 55.0 | 16.0 | 60.0 | 17.0 | 65.0 | 20.0 | 250.4 | 73.5 |
| J182952 | 4.5 | 0.5 | 7.0 | 0.8 | 2.7 | 0.3 | 30.0 | 3.4 | 6.0 | 0.7 | 36.0 | 4.0 | 270.5 | 30.3 |
| J163152 | <0.03 | <0.03 | <6.0 | <6.0 | <5.0 | <5.0 | 18.0 | 18.0 | <8.0 | <8.0 | 10.0 | 10.0 | – | – |
| J162625 | <0.02 | <0.005 | <4.0 | <1.0 | <2.0 | <0.5 | 200.0 | 50.0 | 35.0 | 8.7 | 100.0 | 25.0 | – | – |
| J032838 | 2.0 | 0.09 | <0.07 | <0.03 | 0.8 | 0.04 | 17.0 | 0.8 | 11.0 | 0.5 | 20.0 | 95.0 | 100.2 | 4.7 |
| J032848 | <0.5 | <0.05 | <2.0 | <0.2 | <2.0 | <0.2 | 20.0 | 2.0 | 15.0 | 1.5 | 25.0 | 2.5 | 100.3 | 10.0 |
| J032851 | <0.6 | <0.08 | <2.0 | <0.3 | <1.5 | <0.2 | 40.0 | 5.5 | 30.0 | 4.1 | 20.0 | 2.7 | – | – |
| J032859 | 7.0 | 0.2 | 4.0 | 0.1 | 1.0 | 0.03 | 200.0 | 5.8 | 55.0 | 1.6 | 50.0 | 1.5 | – | – |
| J032911 | 1.8 | 0.1 | <1 | <0.08 | <1.0 | <0.08 | 45.0 | 3.6 | 7.0 | 0.6 | 3.0 | 24.0 | – | – |
| J041858 | <0.3 | <0.009 | <6.0 | <2.0 | <4.0 | <1.0 | 8.0 | 2.5 | 20.0 | 6.2 | 2.5 | 78.0 | – | – |

The uncertainty is estimated to be ~23-32% for the column densities and molecular abundances.

enrichment of abundances of the $H_3^+$ isotopologues at <20–30 K. The initial gas-phase deuterium enrichment of $H_3^+$ is even more pronounced in cold, dense regions, where some destructive neutral species, especially CO, become severely depleted onto dust grain surfaces. Thus, $H_2D^+$ becomes abundant below ~20-30 K, both because of the increased difficulty of overcoming the energy barrier to re-form $H_3^+$ from $H_2D^+$ ($\Delta$E ~ 230 K) and because of the freeze-out of molecules like CO that can destroy $H_2D^+$ (e.g., Roberts & Millar 2000ab). Because of the relatively high abundance of $H_3^+$, $H_2D^+$ is the dominant deuteron donor below ~20-30 K, and is one of the principal ions in cold, dense gas (Roberts & Millar 2000ab). This effect is further enhanced by the reduction in the $H_2$ ortho/para ratio at low temperatures, where the energy difference between the ortho and para ground states ($\Delta$ E ~170 K) could provide the internal energy for the back reaction (e.g., Pagani et al. 2011).

Deuterium fractionation also starts from a limited number of exchange reactions of HD with $CH_3^+$ and $C_2H_2^+$:

$$HD + CH_3^+ \rightarrow CH_2D^+ + H_2 + \Delta E \quad (2)$$

$$HD + C_2H_2^+ \rightarrow C_2HD^+ + H_2 + \Delta E \quad (3)$$

The reactions of HD with $CH_3^+$ and $C_2H_2^+$ are more exothermic than $H_3^+$, with $\Delta$E ranging from 480 K to 660 K for reaction (2) and ~550 K for reaction (3) (Roueff et al. 2013). The reverse reactions with these species occur readily above ~80 K. Hence, the $CH_2D^+$ and $C_2HD^+$ ions survive more easily in warmer gas than $H_2D^+$ and allow deuterium fractionation to proceed effectively at warmer temperatures of ~30-80 K (e.g., Roberts & Millar 2000ab; Roueff et al. 2005; Parise et al. 2009).

There is thus both a cold and a warm deuteration channel. These deuterated ions react further with abundant molecules such as CO and $N_2$, transferring deuterium atoms to new molecules. For the case of Class 0/I proto-BDs, simulations have shown that the kinetic temperature is nearly constant in the cores at ≤10 K, and rises to ~20-30 K only in the innermost (~10-20 au) region close to the jet/outflow launching zone and due to the contribution from the stellar irradiation (Riaz et al. 2019b). We therefore expect the cold deuteration channel to be mainly active in the proto-BDs.

In the following sections, we describe the cold and warm formation pathway for the various deuterated species observed in this work. The molecular abundances in all plots and tables are with respect to $H_2$. The CO abundances have been derived from the $C^{17}O$ abundance (Riaz et al. 2019). This ensures that the CO abundances are not affected by the opacity effects since $C^{17}O$ is an optically thin isotopologue. The CO depletion factor, $f_D$, is defined as $X_{can}$/X, where $X_{can}$ is the canonical abundance determined by Frerking et al. (1982) towards dark cores ($X_{can} = 4.8 \times 10^{-8}$), and the abundance X is the ratio of the $C^{17}O$ and $H_2$ column densities (Bacmann et al. 2002). For $f_D$ <1, CO is expected to be in the gas-phase, while higher values of $f_D$ ~10 indicate a large fraction (>80%) of CO is frozen onto dust grains. Table 5 lists the values of $f_D$ and the D/H ratios derived from the various deuterated species for the sample. In Figs. 1; 3; 4; 5; 6, we have shown the best linear fit to the observed data points. For a quantitative comparison of the correlations between the molecular abundances, $f_D$, and the D/H ratios, we have calculated the Pearson correlation coefficient, *r*, from the best-fit to the observed data points, and the value for *r* is noted in each plot. The Pearson correlation coefficient is a measure of the covariance of the two variables divided by the product of their standard deviations. A value of ±1 for *r* indicates a strong correlation or anti-correlation between the two parameters for a positive or negative slope of the best-fit, respectively.

### 4.2 $DCO^+$

$DCO^+$ is among the primary tracers of deuterium chemistry in the interstellar medium due to its relatively high abundance and accessible rotational transitions (e.g., Bergin et al. 2007). All of the Stage 0/I proto-BDs in our sample, except J163136, show a detection in the $DCO^+$ (3-2) line, as well as the Stage II object J041858. The key gas-phase formation pathway for $DCO^+$ that is active at cold temperatures of <20–30 K is via $H_2D^+$ (e.g., Pagani et al. 2011),

$$CO + H_2D^+ \rightarrow \frac{2}{3}HCO^+ + \frac{1}{3}DCO^+ + \frac{1}{3}H_2 + \frac{2}{3}HD. \quad (4)$$

The warm deuteration channel for the formation of $DCO^+$ is via the reaction:

$$CH_2D^+ + O \rightarrow DCO^+ + H_2, \quad (5)$$

which would make $DCO^+$ abundant at warmer temperatures of ~30-80 K (e.g., Favre et al. 2015).

Figure 1a shows a comparison of the $DCO^+$ and $HCO^+$ abundance vs. the CO abundance for the proto-BDs. The $HCO^+$ abundances are at least a factor of ~2 higher than $DCO^+$. As noted, the formation of these species via reaction (4) would result in 2/3 $HCO^+$ and 1/3 $DCO^+$. Note that the main formation pathway of $HCO^+$ is via reaction with $H_3^+$ (e.g., Riaz et al. 2019), and the higher $HCO^+$ abundances compared to $DCO^+$ do not indicate that both are forming via the same reaction with $H_2D^+$. A slight increase is seen in the $DCO^+$ and $HCO^+$ abundances with [CO], although the large spread results in a weak correlation of ~50%-60%. $DCO^+$ and $HCO^+$ are high-density tracers with a critical density that is ~2 orders of magnitude higher than CO. The lack of a tight correlation could be due to differences in the origin of emission, where CO could be tracing the low-density gas in the outer envelope/outflow regions, while $DCO^+$ and $HCO^+$ trace the high-density inner envelope/pseudo-disk regions (e.g., Riaz et al. 2019).

We have also obtained observations in the $H^{13}CO^+$ (3-2) line for the proto-BDs. Figures 1bc show the trend in the $DCO^+/H^{13}CO^+$ and $DCO^+/HCO^+$ ratios vs. the CO depletion factor. The $DCO^+/H^{13}CO^+$ ratios (~0.3–5) are comparatively higher and show a narrower range than the $DCO^+/HCO^+$ ratios (~0.008–0.6). This indicates that the $HCO^+$ line is likely optically thick, and thus the $DCO^+/HCO^+$ ratios can be considered as lower limits. The optically thin isotopologue of $H^{13}CO^+$ can provide a better assessment of the abundance of the $HCO^+$ molecule (e.g., Riaz et al. 2018); therefore we consider the $DCO^+/H^{13}CO^+$ ratio to be a more reliable measure of the fractionation of $DCO^+$.

Both the $DCO^+/HCO^+$ and $DCO^+/H^{13}CO^+$ ratios

Table 5: The D/H ratios and the CO depletion factors

| Object | [$DCO^+/HCO^+$] | [$DCO^+/H^{13}CO^+$] | [DCN/HCN] | [$DCN/H^{13}CN$] | [DNC/HNC] | [$DNC/HN^{13}C$] | [$N_2D^+/N_2H^+$] | $f_D$ |
|---|---|---|---|---|---|---|---|---|
| J182854 (Stage I) | 0.05 | 0.8 | <0.05 | – | <0.05 | – | <0.02 | 5.6 |
| J182844 (Stage 0) | 0.6 | 2.7 | 0.02 | <0.5 | 0.8 | 2.6 | 0.03 | 4.6 |
| J183002 (Stage 0) | 0.008 | 0.7 | – | – | 0.02 | – | <0.01 | 0.8 |
| J182959 (Stage I) | 0.4 | 4.7 | 0.2 | – | 0.2 | 3.2 | 0.02 | 0.2 |
| J182953 (Stage 0) | 0.008 | 0.3 | <0.3 | – | <0.07 | – | – | 0.8 |
| J182856 (Stage 0) | 0.2 | 2.0 | 0.09 | <1.3 | 0.02 | <0.5 | 0.005 | 2.1 |
| J182952 (Stage 0) | 0.3 | 2.1 | 0.4 | 0.4 | 0.2 | 2.3 | 0.02 | 1.9 |
| J163143 (Stage I) | 0.03 | – | <0.3 | – | <0.1 | – | – | 2.1 |
| J163136 (Stage I) | <0.04 | – | <0.1 | – | <0.1 | – | – | 1.3 |
| J163152 (Stage 0) | 0.04 | – | <0.1 | – | <0.1 | – | – | 0.9 |
| J162625 (Stage 0) | 0.4 | – | 0.2 | – | – | – | – | 2.4 |
| J032838 (Stage I) | 0.4 | 3.1 | <0.03 | <0.4 | 0.05 | 1.3 | <0.008 | 34.3 |
| J032848 (Stage I) | 0.1 | <4.0 | <0.03 | <0.2 | 0.08 | <1.0 | 0.02 | 12.4 |
| J032851 (Stage 0) | 0.04 | <2.4 | – | – | <0.03 | <0.7 | – | 5.4 |
| J032859 (Stage I) | 0.07 | 2.0 | 0.04 | 0.6 | 0.1 | 6.6 | – | 18.5 |
| J032911 (Stage I) | 0.02 | 0.5 | <0.07 | – | <0.4 | 1.2 | – | 2.3 |
| Average (STD) | 0.18 (0.19) | 1.9 (1.3) | 0.16 (0.14) | 0.47 (0.1) | 0.19 (0.25) | 2.9 (1.8) | 0.02 (0.008) | |
| J041858 (Stage II) | 0.48 | – | <0.05 | – | 1.15 | – | – | 19.2 |
| J182940 (Stage I-T) | <0.01 | – | <2.0 | – | <2.0 | – | – | 4.8 |
| J182927 (Stage I-T) | <0.06 | – | <1.0 | – | <2.0 | – | – | 1.4 |
| J182952 (Stage II) | <0.01 | – | <0.2 | – | <0.2 | – | – | 3.4 |

The uncertainty is estimated to be ~23-32% on the D/H ratios and the CO depletion factors. The mean ratios and standard deviation (STD) are calculated only for the Stage 0/I objects, excluding the upper limits. A '–' indicates that observations are not available.

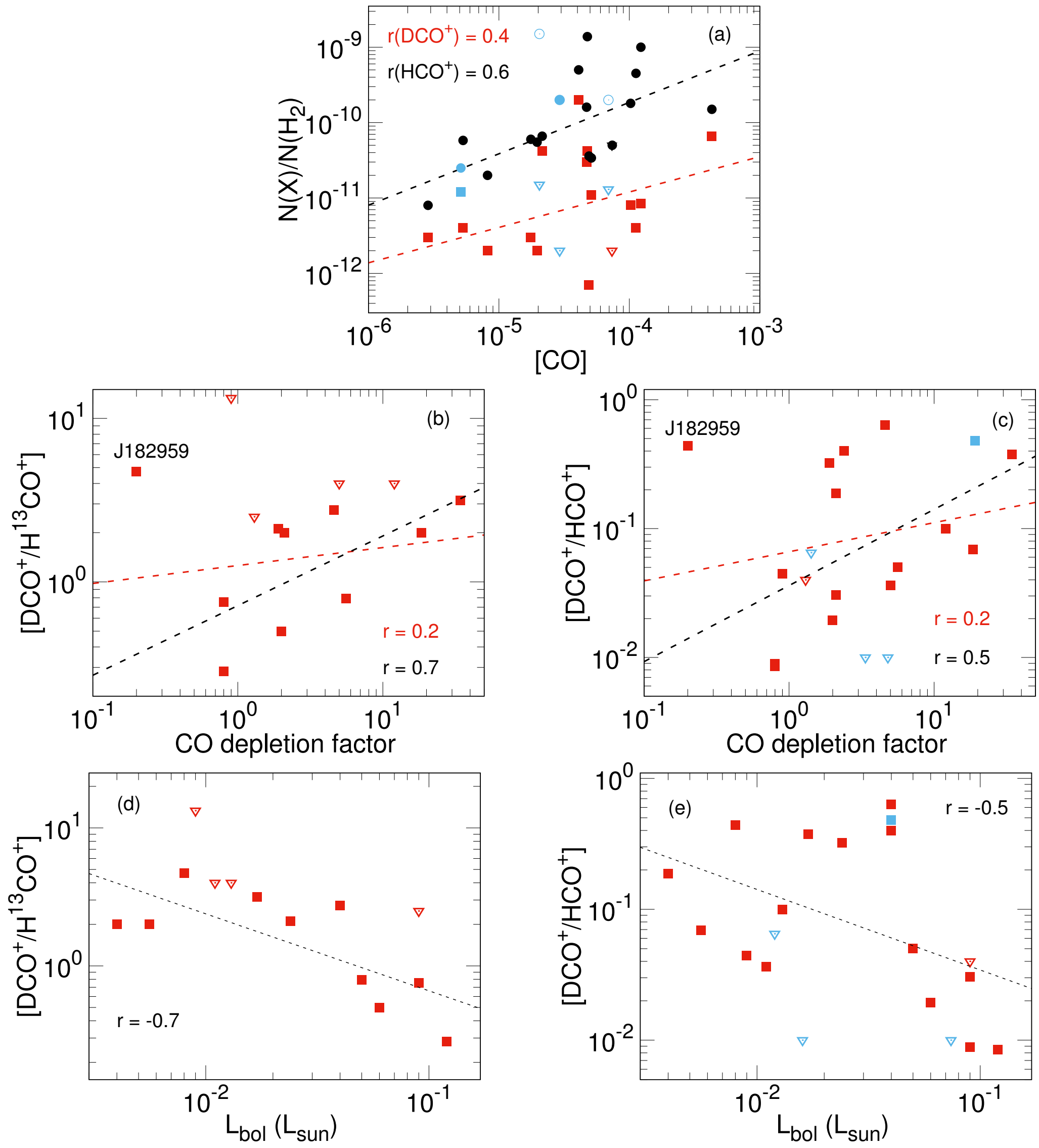

**Figure 1. (a)** Correlations of the $DCO^+$ (red) and $HCO^+$ (black) abundances (with respect to $H_2$) with the CO abundances. **(b,c)** Correlations of the $DCO^+/H^{13}CO^+$ and $DCO^+/HCO^+$ ratios with the CO depletion factor. **(d,e)** Correlations of the $DCO^+/H^{13}CO^+$ and $DCO^+/HCO^+$ ratios with the bolometric luminosities. In all plots, the Stage I-T/II BDs are plotted as blue squares. Diamonds denote the non-detections (upper limits). Dashed lines are the best-fit to the observed data points for the Stage 0/I proto-BDs excluding the upper limits and the data points for Stage I-T/II BDs. The correlation coefficients for the best-fits are noted in the plots. In plots **(b,c)**, red dashed line is the fit to all proto-BDs while black dashed line is the fit excluding J182959.

show a weak correlation with the CO depletion factor ($r$ = 0.2). However, if the one anomalous point of J182959 is excluded from the linear fit then a stronger correlation is seen with $r$ = 0.7 for $DCO^+/H^{13}CO^+$ and $r$ = 0.5 for $DCO^+/HCO^+$. The main production route of $DCO^+$ is the low temperature channel, which requires CO in the gas phase. The increase in $H_2D^+$ formation at low temperatures of <20-30 K is expected to enhance the formation of $DCO^+$. However, CO can remain in the gas phase down to a temperature of ~20 K. At gas temperatures of >20 K, CO can destroy $H_3^+$ and suppress the formation of $H_2D^+$ and hence $DCO^+$. Since CO is also required to form $DCO^+$ via gas-phase reaction (4), the ideal condition is when the gas temperature is just above the CO freeze-out temperature and the CO gas phase abundance is low.

The one anomalous point for J182959, with a high $DCO^+/H^{13}CO^+$ ratio of ~4.5 for a low CO depletion $f_D$ ~0.2, suggests $DCO^+$ formation via the warm channel in this proto-BD. A gas kinetic temperature of ~20-30 K can only be reached in a proto-BD close to the jet/outflow launching regions (<10 au), due to strong accretion and outflow activity with sudden accretion bursts or episodic jet ejections. J182959 shows strong CN emission but no HCN detection, which is also indicative of strong accretion activity (Riaz et al. 2018). If the formation pathway of $DCO^+$ in this proto-BD is via the warm pathway then some form of thermal or non-thermal process is required that raises the local gas temperature to higher than ~20 K.

Figure 1d shows a rise in the $DCO^+/H^{13}CO^+$ ratio with decreasing $L_{bol}$ for the proto-BDs ($r$ = -0.7). The low-est luminosity proto-BDs are expected to be cooler than the higher $L_{bol}$ objects, and it is likely that these cores provide the ideal environment for enhanced formation of $DCO^+$. The central object mass in the lowest luminosity cases would be <10 $M_{Jup}$, and thus desorption due to thermal radiation of the central proto-BD would also be weak. On the other hand, the large spread of about two orders of magnitude seen in the $DCO^+/HCO^+$ ratios (Fig. 1e) is notable and suggests that other factors such as high opacity and/or inclination effects could produce a large scatter at a given $L_{bol}$. The wide range in the ratios also hints at warm+cold formation of $DCO^+$, since the warmer the dust gets, the more the ratios diverge. There is also no clear correlation seen between the $DCO^+/H^{13}CO^+$ or $DCO^+/HCO^+$ ratios and the evolutionary stage of the system; for e.g., J182844 and J183002 show the highest and lowest $DCO^+/HCO^+$ ratios, respectively, and are both Stage 0 objects (Table 5). The large scatter could be due to differences in the physical+chemical structure, and some proto-BD cores may be more chemically evolved than others, irrespective of the evolutionary stage or the bolometric luminosity.

### 4.3 DCN and DNC

DCN has a critical density an order of magnitude higher than other deuterium species (Table 2), and thus probes the densest regions deep inside the proto-BD core. The DCN (3-2) line is detected in 9/16 Stage 0/I proto-BDs. None of the Stage I-T/II objects show DCN emission. The DNC (3-2) line is detected in 8/16 Stage 0/I proto-BDs, with weak emission detected at a ~2-σ level in the Stage II object J041858. There are only two objects with $H^{13}CN$ detection and 5 with $HN^{13}C$ detection. A similar range is found in the DCN/HCN (0.02-0.4), DNC/HNC (0.02-0.8), and $DCN/H^{13}CN$ (0.4-0.6) ratios, while $DNC/HN^{13}C$ (1.3-6.6) are higher by a factor of ~2-16.

The gas-phase formation pathway for DCN at cold (<20-30 K) temperatures (e.g., Roueff et al. 2005) is proposed as:

$$H_2D^+ + CO \rightarrow H_2 + DCO^+ \tag{6}$$

$$DCO^+ + HNC \rightarrow DCNH^+ + CO \tag{7}$$

$$DCNH^+ + e^- \rightarrow DCN + H \tag{8}$$

The cold gas-phase formation pathway for DNC is proposed as:

$$H_2D^+ + CO \rightarrow H_2 + DCO^+ \tag{9}$$

$$DCO^+ + HCN \rightarrow HCND^+ + CO \tag{10}$$

$$HCND^+ + e^- \rightarrow DNC + H \tag{11}$$

The formation via warm (~30-80 K) channel (e.g., Roueff et al. 2005; 2007) for DCN is proposed as:

$$CH_2D^+ + H_2 \rightarrow CH_4D^+ + h\nu \tag{12}$$

$$CH_4D^+ + e^- \rightarrow CHD + H_2 + H \tag{13}$$

$$CHD + N \rightarrow DCN + H \tag{14}$$

The other possible warm channels for DCN formation are:

$$CH_2D^+ + e^- \rightarrow CHD + H \tag{15}$$

$$CHD + N \rightarrow DCN + H \tag{16}$$

$$CH_2D^+ + N \rightarrow DCN^+ + H_2 \tag{17}$$

$$DNC^+ + H_2 \rightarrow DCNH^+ + H \tag{18}$$

$$DCNH^+ + e^- \rightarrow DCN + H \tag{19}$$

The warm pathway for DNC is proposed as (e.g., Roueff et al. 2007):

$$CH_2D^+ + N \rightarrow DNC^+ + H_2 \tag{20}$$

$$DNC^+ + H_2 \rightarrow DNCH^+ + H, DHNC^+ + H \tag{21}$$

$$DNCH^+ + e^- \rightarrow DNC + H \tag{22}$$

$$DHNC^+ + e^- \rightarrow DNC + H \tag{23}$$

The formation reactions for DNC and DCN suggest that their abundances should increase with $DCO^+$ abundance. Figure 2a shows a strong correlation ($r$ = 0.85-0.87) between the $DCO^+$, DCN, and DNC abundances. We see a potential trend of increasing ratio consistent with the proposed formation pathway.

Fractionation of these neutral molecules can also occur via grain surface chemistry. For e.g., DCN and DNC can form on grains from the freeze-out and deuteration of CN and CD molecules via grain surface (labelled ':gr') reactions, as proposed in e.g., Albertsson et al. (2013):

$$CN : gr + D : gr \rightarrow DCN : gr \tag{24}$$

$$CD : gr + N : gr \rightarrow DCN : gr \tag{25}$$

$$ND : gr + C : gr \rightarrow DNC : gr \tag{26}$$

$$CN : gr + D : gr \rightarrow DNC : gr \tag{27}$$

Such grain surface reactions are enhanced when density increases and heavy species are adsorbed onto grain surfaces (e.g., Aikawa et al. 2012). Thus, the formation of these molecules via surface reactions is expected to be more efficient in the dense inner regions of the proto-BDs where CO is depleted.

Figures 3ab compare the $DCO^+$, DCN, and DNC abundances vs. the CO abundance for the proto-BDs. The DCN abundances are comparatively lower than $DCO^+$, while DNC abundances lie in a similar range as $DCO^+$. As also seen for $DCO^+$, the wide spread in the DCN and DNC abundances could be explained by the difference in the origin of the line emission compared to CO, i.e., DCN and DNC trace the dense regions in the inner pseudo-disk/envelope, while CO traces the low-density outflowing gas and the outer envelope regions.

The DCN/HCN ratios in Fig. 3c show a relatively strong anti-correlation with the CO depletion factor ($r$ = -0.6). Also note that there are no DCN detections for objects with low CO depletion factor of <1 i.e. when CO is in the gas phase. This suggests that the deuteration of HCN becomes active once CO is frozen onto dust grains. The lower DCN/HCN ratios for higher depletion factors of ≥10 can be explained by the difference in the peak emitting regions of these molecules. The DCN emission in a proto-BD peaks in the inner pseudo-disk region at ~100-150 au but shows a sharp drop at radii < 50 au, as predicted by physical+chemical models (Riaz et al. in prep). CO shows a similar intensity profile, and both of these molecules have a central depletion zone (< 50 au). In contrast, HCN peaks in this inner depletion zone at radii < 50 au, while the HCN intensity drops at > 50 au. Thus the DCN/HCN ratio will increase as we move to radii > 50 au or as the CO depletion factor decreases. DCN is still forming via the cold pathway but the clearly opposite peak emitting regions of DCN and HCN result in an anti-correlation with CO. A different trend is seen between the DCN/$H^{13}CN$ ratios and the CO depletion factor (Fig. 3d), where the ratios show a slight rise with stronger CO depletion. However, with only two proto-BDs with $H^{13}CN$ detections, it is difficult to confirm this trend.

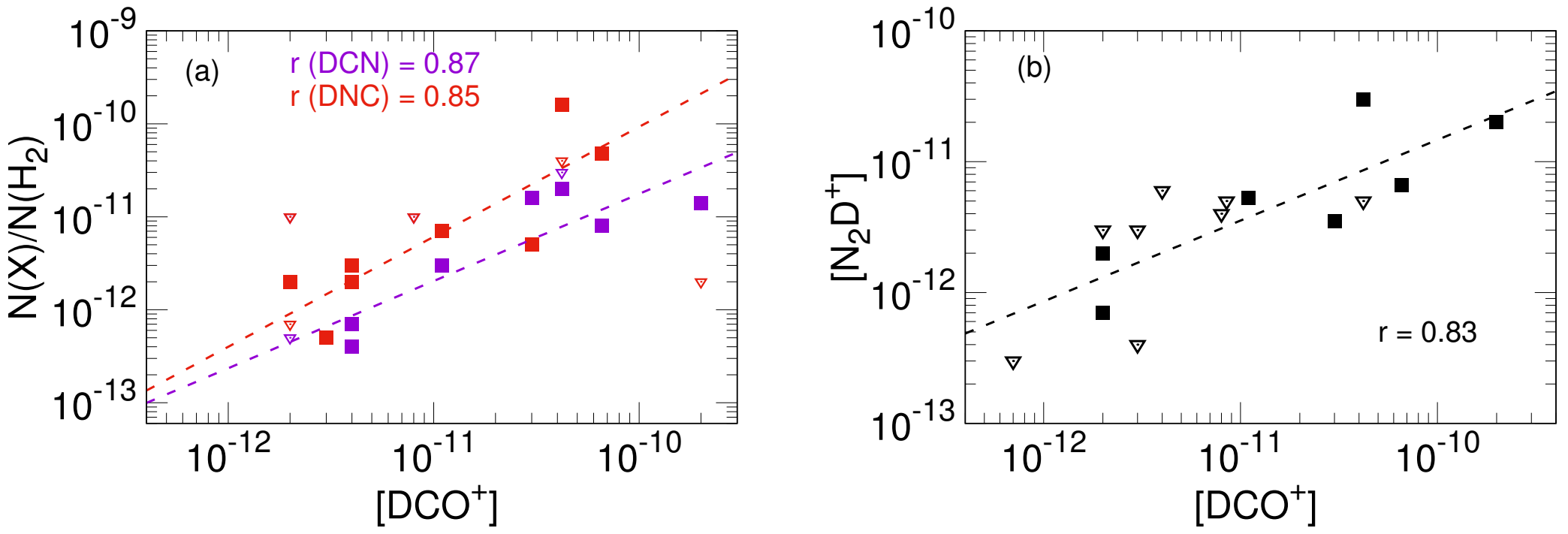

**Figure 2.** A comparison of the $DCO^+$ abundances with DCN and DNC abundances (left) and the $N_2D^+$ abundances (right) for the proto-BDs. Upper limits are denoted by diamonds. Dashed lines are the best-fit to the observed data points excluding the upper limits. The correlation coefficients from the best-fits are noted in the plots.

The DNC/HNC and DNC/$HN^{13}C$ ratios, however, do not show any dependence on the CO abundance (Fig. 3ef). This suggests that the deuteration of HNC is enhanced in the dense inner regions where CO is frozen onto dust grains, and DNC formation in proto-BDs is via the cold channel. Riaz et al. (2018) had noted that HNC shows a nearly flat distribution with CO ($r$ = 0.2) indicating that HNC can avoid depletion in the regions where CO is depleted. If HNC is being formed in proto-BDs through ice surface reactions with $N_2$ ice at very low temperature then its abundance will remain large even in the inner, dense regions where CO is frozen.

In Figs. 4, no strong dependence is seen between the DCN/HCN and DNC/HNC ratios with the bolometric luminosity for the proto-BDs ($r$ = 0.1). The DNC/$HN^{13}C$ ratios show an anti-correlation with $L_{bol}$ ($r$ = -0.6), although there are only 5 proto-BDs with $HN^{13}C$ detection. No dependence is seen between these ratios and the evolutionary stage of the system; the highest and the lowest ratios are found for J162625 and J182844, both of which are Stage 0 objects.

### 4.4 $N_2D^+$

$N_2D^+$ also forms mainly through the low temperature deuteration channel via ion-molecule reaction (Delgarno & Lepp 1984),

$$N_2 + H_2D^+ \rightarrow \frac{2}{3}N_2H^+ + \frac{1}{3}N_2D^+ + \frac{1}{3}H_2 + \frac{2}{3}HD \qquad (28)$$

The formation of $N_2D^+$ is expected to be enhanced at the same low temperature of <20-30 K as $H_2D^+$. However, CO can also destroy $N_2D^+$ and enhance the formation of $DCO^+$ via the reaction,

$$N_2D^+ + CO \rightarrow DCO^+ + N_2, \qquad (29)$$

and the formation of $N_2D^+$ will therefore be aided by a decrease in the CO abundance.

Figure 5a shows a weak correlation between $N_2D^+$ and CO abundances ($r$ ~0.3). At high CO abundance of $>10^{-6}$, $N_2D^+$ is destroyed to form $DCO^+$ via reaction (29), which also results in enhancing the $N_2$ gas density. Chemical models predict that the deuteration capabilities of $DCO^+$ are expected to reach a saturation level for [CO]~$10^{-5}$–$10^{-4}$, at which point the $N_2D^+$ abundance is expected to be 1-2 orders of magnitude lower than [$DCO^+$] (e.g., Pagani et al. 2011). In contrast, we find a similar large spread of over two orders of magnitude in the $DCO^+$ and $N_2D^+$ abundances for the proto-BDs (Fig. 5). This can be explained by a high $N_2$ gas density that would enhance $N_2D^+$ formation via the reaction (28). The lower the binding energy of a species, the lower the desorption temperature at a given gas pressure (or density). The adsorption or binding energy for $N_2$ and CO are 990 K and 1100 K, respectively (e.g., Penteado et al. 2017). This translates into a lower sublimation temperature of ~16 K for $N_2$ compared to ~20 K CO (e.g., Öberg et al. 2010). Due to this, $N_2$ is depleted later than CO, or $N_2$ can survive in the gas phase for a longer time than CO. A balance between low temperature, high $N_2$ gas density and low CO gas phase abundance can enhance both $DCO^+$ and $N_2D^+$ formation in the gas phase. This is evidenced from the strong correlation ($r$ = 0.83) seen between the $DCO^+$ and $N_2D^+$ abundances in Fig. 2b. This could result in similar $DCO^+$ and $N_2D^+$ abundances and a lack of correlation with CO, as observed for the case of proto-BDs.

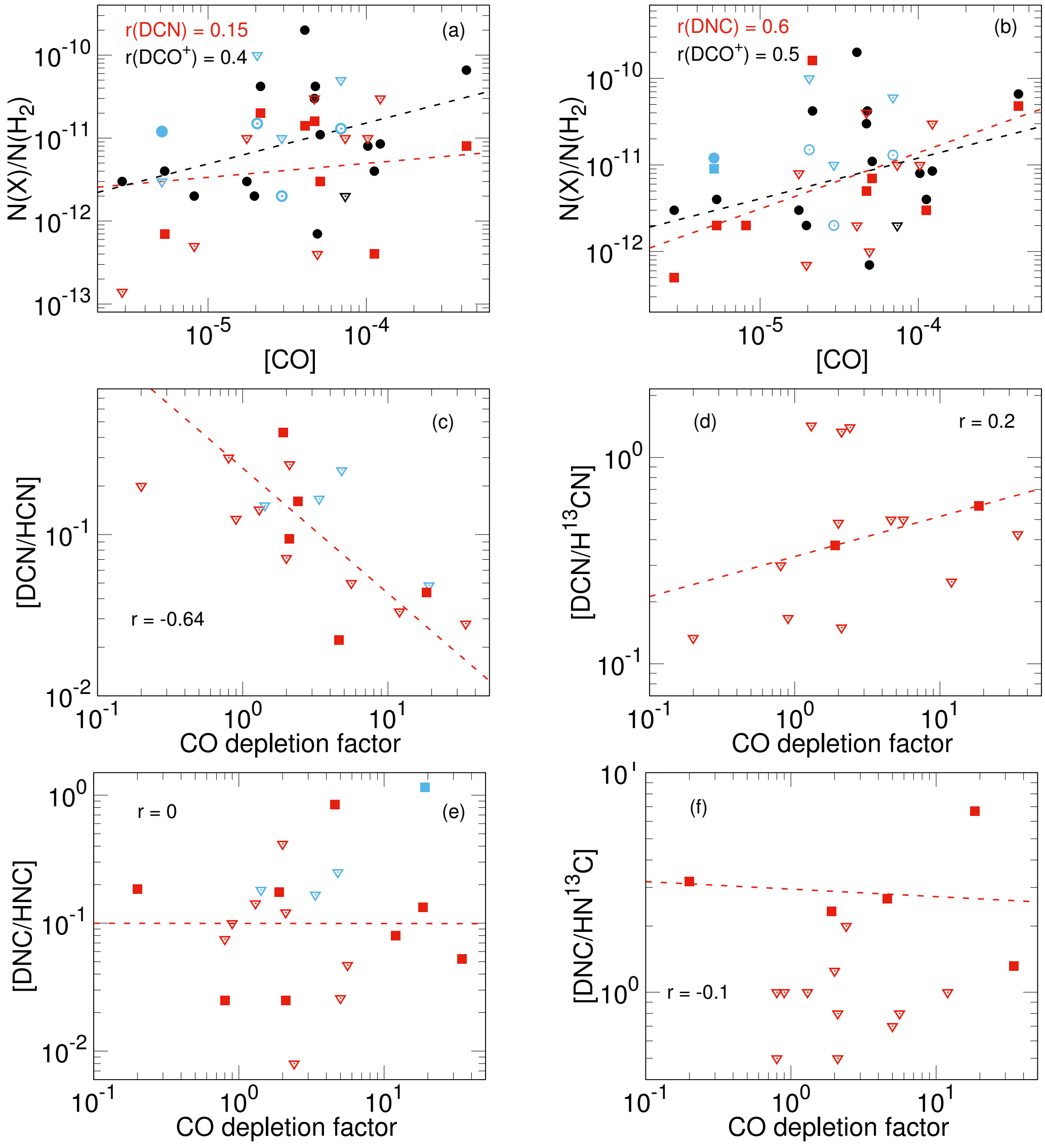

**Figure 3.** **(a)** Correlations of the DCN (red) and $DCO^+$ (black) abundances (with respect to $H_2$) with the CO abundance. **(b)** Correlations of the DNC (red) and $DCO^+$ (black) abundances (with respect to $H_2$) with the CO abundance. **(c)** Correlations of the DCN/HCN ratios with the CO depletion factor. **(d)** Correlations of the $DCN/H^{13}CN$ ratios with the CO depletion factor. **(e)** Correlations of the DNC/HNC ratios with the CO depletion factor. **(f)** Correlations of the $DNC/HN^{13}C$ ratios with the CO depletion factor. In all plots, the Stage I-T/II BDs are plotted as blue squares. Diamonds denote the non-detections (upper limits). Dashed lines are the best-fit to the observed data points for the Stage 0/I proto-BDs excluding the upper limits and the data points for Stage I-T/II BDs. The correlation coefficients for the best-fits are noted in the plots.

### 4.5 $N_2H^+$

$N_2H^+$ follows the same trend as $HCO^+$ but is always under-abundant than $HCO^+$ at any layer or temperature. The main formation pathway of $N_2H^+$ is via the reaction:

$$H_3^+ + N_2 \rightarrow N_2H^+ + H_2, \quad (30)$$

and destroyed via dissociative recombination

$$N_2H^+ + e^- \rightarrow N_2 + H \quad (31)$$

$N_2H^+$ is depleted in the outer envelope region (via reaction 31) where the electron density is high. In deeper regions, $N_2$ is abundant (like CO), due to which $N_2H^+$ production increases via reaction (30) and $N_2H^+$ should be as abundant as $HCO^+$. However, in these deeper layers, there is also a destructive path for $N_2H^+$ via CO

$$N_2H^+ + CO \rightarrow HCO^+ + N_2 \quad (32)$$

This results in maintaining the $HCO^+$ abundance and the correlation between $HCO^+$ and CO. In Fig. 5b, the $N_2H^+$ abundances are comparable to the $HCO^+$ abundances for the proto-BDs, with a similar correlation coefficient with the CO abundance. There may be a balance between formation and destruction of $N_2H^+$ resulting in a constant abundance in the deeper layers that is comparable to $HCO^+$. In Fig. 5c, the $N_2H^+$ abundances are at least a factor of

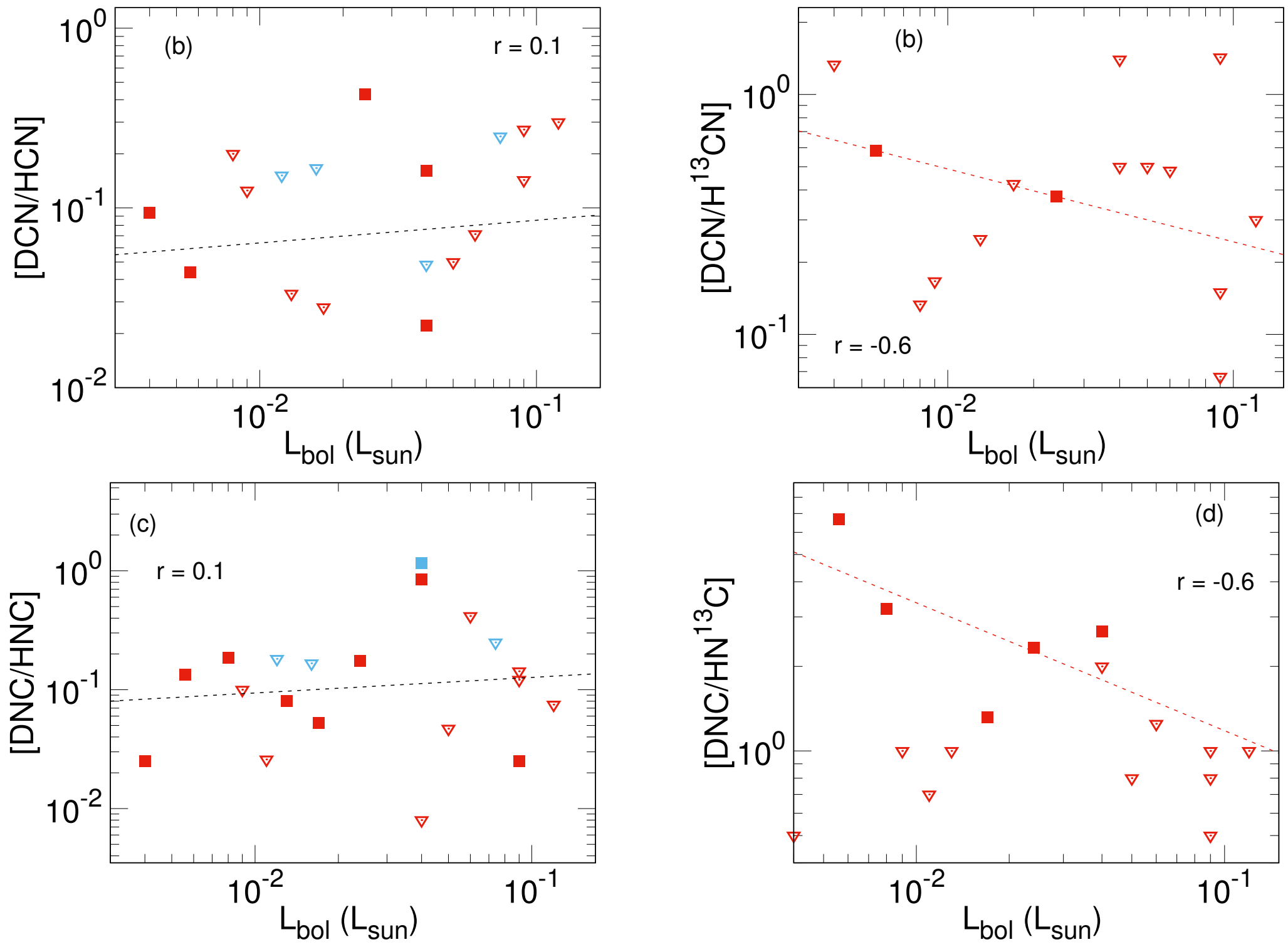

**Figure 4. (a)** Correlations of the DCN/HCN ratios with the bolometric luminosity for the proto-BDs. **(b)** Correlations of the DCN/$H^{13}CN$ ratios with the bolometric luminosity for the proto-BDs. **(c)** Correlations of the DNC/HNC ratios with the bolometric luminosity for the proto-BDs. **(d)** Correlations of the DNC/$HN^{13}C$ ratios with the bolometric luminosity for the proto-BDs. In all plots, the Stage I-T/II BDs are plotted as blue squares. Diamonds denote the non-detections (upper limits). Dashed lines are the best-fit to the observed data points for the Stage 0/I proto-BDs excluding the upper limits and the data points for Stage I-T/II BDs. The correlation coefficients for the best-fits are noted in the plots.

~2 higher than $N_2D^+$ (Fig. 5c). As noted, the formation of these species via reaction (28) would result in 2/3 $N_2H^+$ and 1/3 $N_2D^+$. The main formation pathway of $N_2H^+$ is via reaction (30) with $H_3^+$, which can enhance the $N_2H^+$ abundance compared to $N_2D^+$. Both $N_2H^+$ and $N_2D^+$ show a slight decline in abundances with the CO depletion factor ($r$ = 0.3-0.6; Fig. 5c). A high $N_2$ density can result in enhanced formation of these species despite the destructive pathways via CO (reactions 32; 29).

### 4.6 $N_2D^+/N_2H^+$ ratio

No correlation is seen between the $N_2D^+/N_2H^+$ ratio and the CO depletion factor ($r$ = 0.08; Fig. 5d). This suggests that the deuteration of $N_2H^+$ is also active in the inner, dense regions where CO is severely depleted. The peak emitting region of $N_2D^+$ is at radii <50 au (Riaz et al. in prep). This explains the high $N_2D^+/N_2H^+$ ratios for a high CO depletion factor. Both $N_2H^+$ and $N_2D^+$ are destroyed by electrons and CO, so the $N_2D^+/N_2H^+$ ratio should stay the same. $N_2D^+$ abundance should correlate with $DCO^+$ as shown in the Fig. 2b because both $DCO^+$ and $N_2D^+$ come from $CO/N_2$ reaction with $H_2D^+$.

On the other hand, a clear dependence is seen of the $N_2D^+/N_2H^+$ ratio on the bolometric luminosity of the proto-BD, with a correlation coefficient of 0.6 (Fig. 5e). While there are few data points, the proto-BD with the lowest $L_{bol}$ shows the least $N_2H^+$ deuteration ($N_2D^+/N_2H^+$ = 0.005) and the one with the highest $L_{bol}$ shows a ~10 times higher ratio.

The linear correlation is likely due to shrinking of the peak emitting region of $N_2D^+$; the peak regions for this species is <50 au, and the lowest luminosity objects are predicted to have smaller inner regions (e.g., Riaz et al. in prep). The decline in $N_2D^+/N_2H^+$ ratio with lower $L_{bol}$ probably reflects the physical size of the emitting region rather than differences in the formation pathway. A larger sample combined with physical+chemical modelling is needed to confirm this tentative trend.

No dependence is seen of the $N_2D^+/N_2H^+$ ratio on the evolutionary stage of the system. The highest and lowest ratios are measured for J182844 and J182856, respectively, both of which are Stage 0 objects. We expect that the $N_2D^+/N_2H^+$ ratio should decrease during protostellar evolution due to the internal heating of the core. As the core environment warms up, $H_2D^+$ and its more highly deuterated isotopologues are destroyed. Consequently, all other deuterated molecules formed by reactions with $H_2D^+$ (such as $N_2D^+$) are destroyed as well. The non-detection of $N_2D^+$ in all of the Stage I-T/II objects suggests that desorption due to thermal heating occurs at a later stage in transition from Stage 0/I –> Stage I-T/II, rather than from Stage 0 –> Stage I phase in brown dwarfs.

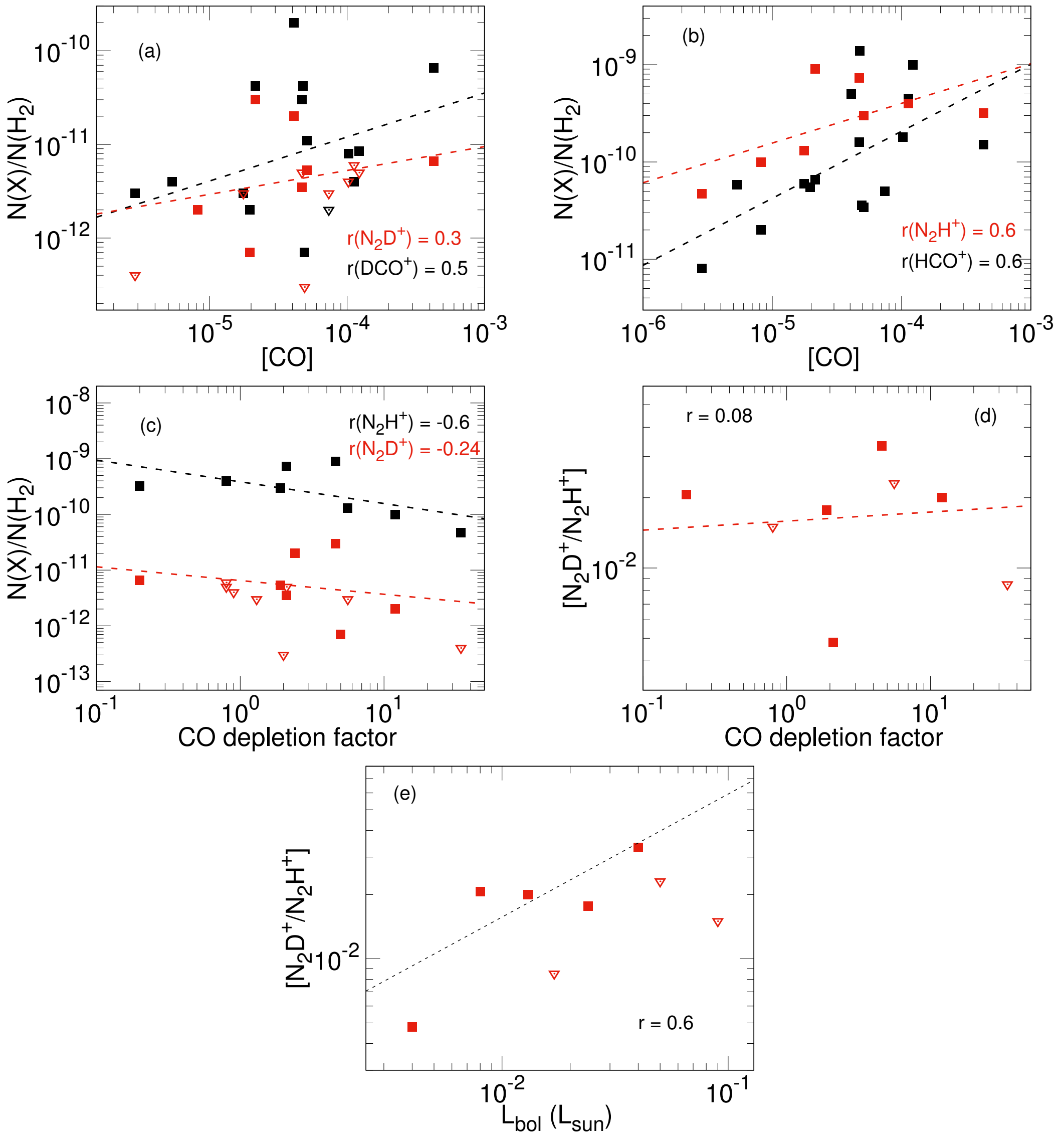

**Figure 5.** **(a)** Correlations of the $N_2D^+$ (red) and $DCO^+$ (black) abundances (with respect to $H_2$) with the CO abundance. **(b)** Correlations of the $N_2H^+$ (red) and $HCO^+$ (black) abundances (with respect to $H_2$) with the CO abundance. **(c)** Correlations of the $N_2D^+$ (red) and $N_2H^+$ (black) abundances (with respect to $H_2$) with the CO depletion factor. **(d)** Correlations of the $N_2D^+/N_2H^+$ ratios with the CO depletion factor. **(e)** Correlations of the $N_2D^+/N_2H^+$ ratios with the bolometric luminosity. Diamonds denote the non-detections (upper limits). Dashed lines are the best-fit to the observed data points for the Stage 0/I proto-BDs excluding the upper limits. The correlation coefficients for the best-fits are noted in the plots.

## 5 DISCUSSION

### 5.1 Comparison with low-mass protostars

Figure 6 probes the trends in the D/H ratios derived using various deuterated species between Class 0/I proto-BDs and low-mass Class 0/I protostars. The data for the protostars has been compiled from single-dish observations presented in Hirota et al. (2001; 2003), Jørgensen et al. (2004), and Emprechtinger et al. (2009). The Hirota et al. (2001; 2003) data is from NRO and NRAO observations in the $J$ =1-0 and 2-1 lines of $DCO^+$, DNC, $HN^{13}C$, $H^{13}CO^+$. The column densities were derived using the LVG analysis using an excitation temperature of 10 K and a distance of 140-145 pc for the Taurus and Ophiuchus protostars. The Jørgensen et al. (2004) data is from JCMT observations in the $J$ = 3-2 lines of $DCO^+$ and DCN. The Ratran radiative transfer code was used to calculate the column densities. The distance is 220 pc for Perseus, 140 pc for Taurus, and 160 pc for Ophiuchus. The Emprechtinger et al. (2009) data is from IRAM 30m observations in the $J$ = 1-0, 2-1, 3-2 lines of $N_2D^+$ and $N_2H^+$. The column densities and abundances were derived using the CTEX method using an excitation temperature of 8-10 K and a distance of 220 pc for Perseus. The $H_2$ column densities in Jørgensen et al. (2004) and Emprechtinger et al. (2009) have been derived from JCMT continuum observations using the same method as used in this work. Our search did not lead to finding the $DCO^+/H^{13}CO^+$ and $DCN/H^{13}CN$ ratios for a large number of low-mass protostars, due to which we could not compare those ratios with proto-BDs. While the excita-tion temperature considered in these studies is similar to our assumed value of 10 K, the main difference is in the distance to Perseus. We have used the latest estimate from Gaia of ~292 pc while previous works have used a distance of 220 pc. A smaller distance of 220 pc would result in a higher total mass and the $H_2$ number and column density by a factor of ~1.7. The resulting molecular abundances would be lower by a factor of ~1.7. The resulting D/H ratios and the trends seen in Fig. 6, however, would be the same.

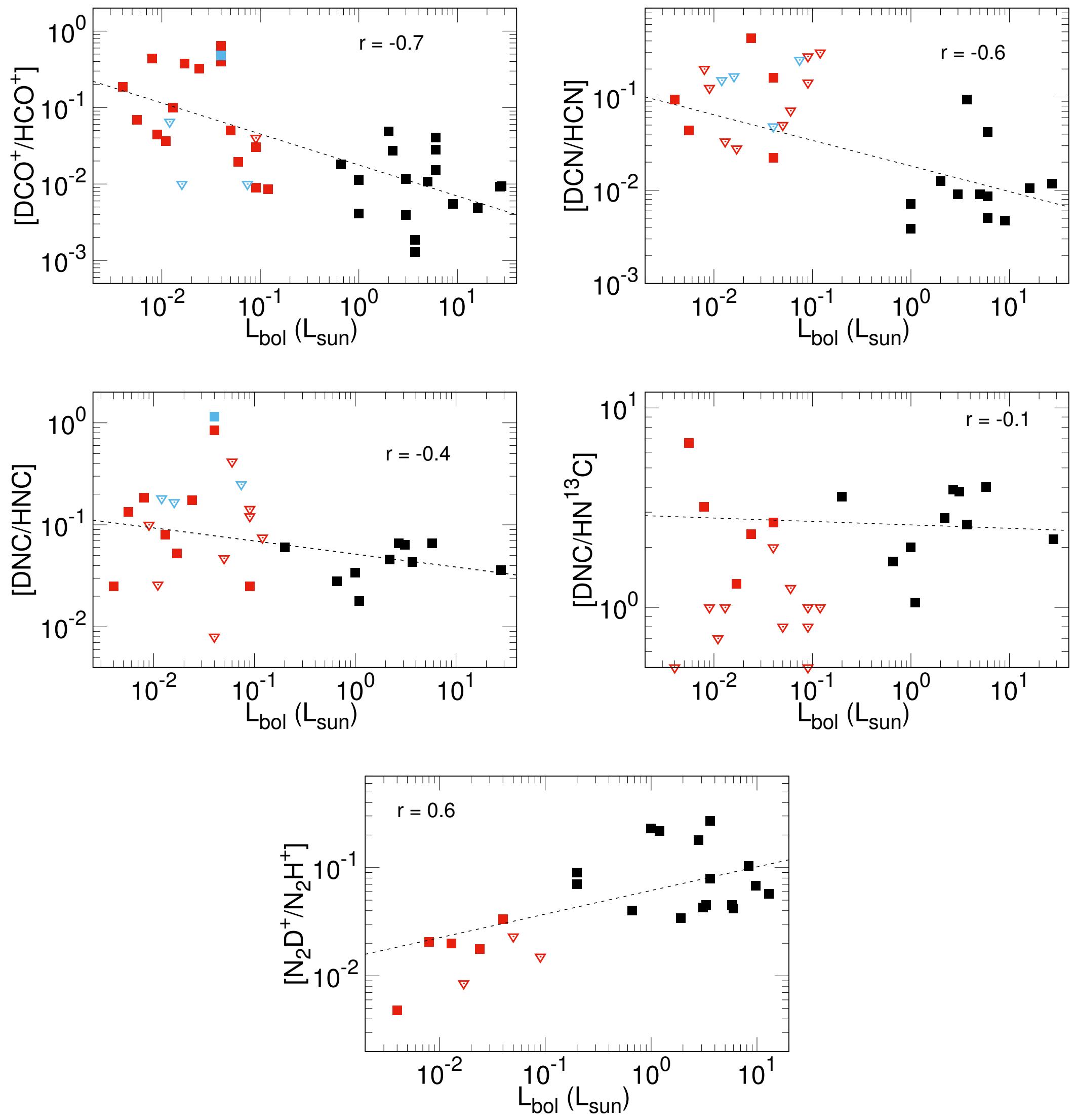

**Figure 6.** Correlations between the D/H ratios for the proto-BDs (red) and low-mass protostars (black) derived using various deuterium species. The data for the protostars has been compiled from the works of Hirota et al. (2001; 2003), Jørgensen et al. (2004), and Emprechtinger et al. (2009). In all plots, the Stage I-T/II BDs are plotted as blue squares. Diamonds denote the non-detections (upper limits). Dashed lines are the best-fit to the observed data points for the Stage 0/I proto-BDs excluding the upper limits and the data points for Stage I-T/II BDs. The correlation coefficients for the best-fits are noted in the plots.

Over a wide range in the bolometric luminosities spanning ~0.002–40 $L_\odot$, we find a trend of higher $DCO^+/HCO^+$ (r = -0.7) and DCN/HCN (r = -0.6) ratios, nearly constant DNC/HNC (r = -0.4) and $DNC/HN^{13}C$ (r = -0.1) ratios, lower $N_2D^+/N_2H^+$ ratios (r = 0.6) in the proto-BDs compared to protostars (Fig. 6). Deuterium fractionation of various molecules significantly varies from core to core and even within each core, due to the differences in the kinematical age, degree of depletion, ionization fraction, chemical evolution, evolutionary stages, and/or the inclination angle of the system. The combination of these effects can produce the large spread seen in the D/H ratios.

### 5.2 Comparison with other astronomical objects

We have also conducted a wider comparison of the D/H ratios in proto-BDs with solar system planets, comets, chondrites, cold molecular clouds ($n_{H_2}$ = 3×$10^4$ $cm^{-3}$, T = 10 K), hot cores ($n_{H_2}$ = $10^7$ $cm^{-3}$, T > 80 K), and low-mass protostars, as shown in Fig. 7. The D/H ratios in Fig. 7 are indicative of the evolutionary physics behind these various objects, and can be used to illustrate evolutionary stages and various fractionation mechanisms. Also plotted are the ele-mental primordial and the proto-solar D/H ratios. The D/H ratios in the ISM and the giant planets (Jupiter, Saturn, Neptune, Uranus) are the molecular ratios measured from the $HD/H_2$ species, due to which these are close to the primordial elemental value. The D/H ratio in the solar nebula ($2.5\pm0.5 \times10^{-5}$) is estimated from the Jovian and Saturnian D/H ratios (e.g., Lellouch et al. 1996; Griffin et al. 1996; Feuchtgruber et al. 1997). The molecular D/H ratios for the terrestrial planets (Earth, Venus, Mars), comets and chondrites are measured from the $HDO/H_2O$ species (Donahue & Pollack 1983; Owen et al. 1998; Lécuyer et al. 1999; Meier et al. 1998; Hartogh et al. 2011; Cleeves et al. 2014).

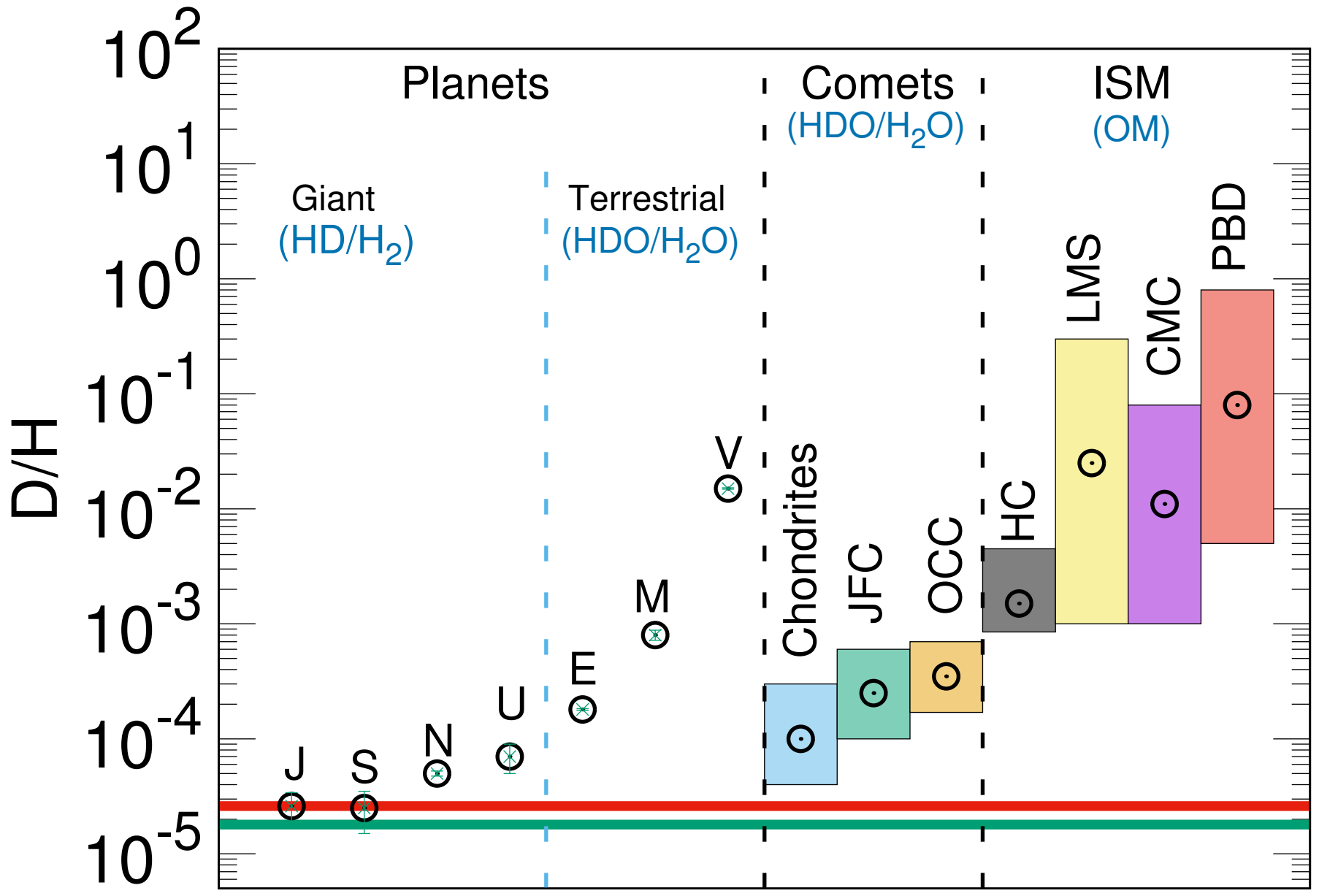

**Figure 7.** The D/H ratios measured for the proto-BDs compared to various objects. The labels are: Jupiter (J), Saturn (S), Neptune (N), Uranus (U), Earth (E), Mars (M), Venus (V), Jupiter Family Comets (JFC), Oort Cloud Comets (OCC), Hot Cores (HC), low-mass protostars (LMS), cold molecular clouds (CC), proto-BDs (PBD). The box length shows the full range in the measurements. Red and green lines mark the primordial and ISM ratios, respectively. The organic molecules ('OM') and other molecular species used to derive the D/H ratios are discussed in the text.

We have plotted the full range in the $DCO^+/HCO^+$, DCN/HCN, DNC/HNC, and $N_2D^+/N_2H^+$ ratios for proto-BDs and low-mass protostars (Tables 5). The ratios for hot cores and cold molecular clouds have been compiled from the works of Robert et al. (2000), Gensheimer et al. (1996), Brown & Millar (1989), Texeira et al. (1999). The D/H ratios in the hot cores are derived from HCN, $CH_3CN$, and $NH_3$ molecules, while the ratios in the cold molecular clouds are derived from the $HCO^+$, $NH_3$, HCN, $H_2CO$, $HC_3N$, $C_3H_2$, $C_2H$, and $HC_5N$ molecules.

The high D/H ratios for low-mass protostars, cold molecular clouds, and proto-BDs is a sign that fractionation mechanisms are efficient in these cold, dense objects. Hot cores are warm and dense that result from the formation of young massive stars which heat the surrounding interstellar medium. High-mass protostars tend to form in warmer environments than their low-mass counterparts. Thus their D/H ratios are lower than their cold interstellar counterparts. Low-mass protostars share ratios with both hot cores and cold molecular clouds, suggesting that both cold and warm deuteration channels could be active in these objects. The D/H ratios for the proto-BDs show the most overlap with the measurements in low-mass protostars and cold molecular clouds. This is expected given the very dense and cold ($n_{H_2}$ $\geq10^6$ cm$^{-3}$, T $\leq$10 K) interior of the proto-BDs, providing the ripe conditions for enhanced deuterium fractionation in these cores.

### 5.3 Comparison with Chemical Models

Several chemical models have calculated the chemical composition of grain mantles using gas phase and grain surface reactions, in order to determine the concentration of deuterated molecules relative to their hydrogenated counterparts in grain mantles. Some models include depletion of gas-phase species in an active sense (e.g., Roberts et al. 2002), while others include surface fractionation processes (e.g., Aikawa et al. 2005). Fractionation on grain surfaces is thought to occur following the formation of a large atomic D/H ratio in the gas. The deuterium and hydrogen atoms, after landing on surfaces, then react with a variety of heavy species leading to both deuterated and normal isotopologues (e.g., Tielens 1983).

In order to achieve the high D/H ratios as measured here for the proto-BDs, very high atomic D abundances are required during the period when ice mantles can accumulate with D/H >0.01. An important route to high levels of deuteration is via reactions between species accreted onto the surfaces of dust grains (grain surface reactions), which can lead to high molecular D/H ratios in cold molecular clouds and pre-stellar cores (e.g., Turner 2001; Robert &

Millar 2002). The models by Robert & Millar (2002; 2003) have also shown that the inclusion of $H_2D^+$ and $D_3^+$ can drive up the atomic D/H ratio to a level where grain surface reactions can account for the high abundances of deuterated species. This effect is greatest in regions where molecules are heavily depleted by freeze-out onto dust grains. Multi-deuterated species such as $D_3^+$ can propagate D to other species through gas-phase ion-molecule reactions more efficiently than mono-deuterated species such as $H_2D^+$ (e.g., Aikawa et al. 2012). The high D/H ratio also propagates to the grain surface. Deuterated $H_3^+$ dissociatively recombines with electrons to produce D atoms, which are then adsorbed onto grains and deuterate the ice mantle species.

The high D/H ratios measured for the proto-BDs suggest that both low-temperature <20 K gas-phase ion-molecule deuteron transfer reactions, likely including multi-deuterated species like $D_3^+$, as well as grain surface reactions can result in enhanced deuterium fractionation in these objects. For a more direct comparison, we consider the gas-grain chemical model by Albertsson et al. (2013), which includes both gas-phase and surface species, and an extended multi-deuterated chemical network. These models predict $DCO^+$/$HCO^+$ and $N_2D^+$/$N_2H^+$ ratios in the range of ~0.01-1, and gas-phase DCN/HCN and DNC/HNC ratios in the range of ~0.001-0.1. In comparison, the $DCO^+$/$HCO^+$, $N_2D^+$/$N_2H^+$, DCN/HCN and DNC/HNC ratios for the proto-BDs are within the range predicted by these models. Large enhancements of the molecular D/H ratios as observed here for proto-BDs provide evidence that accretion (grain surface chemistry) may be occurring for the neutral molecules of DCN and DNC. Given the high densities and cold temperatures found in proto-BDs, molecules that hit a grain are likely to stick to it efficiently, thus resulting in enhanced fractionation of deuterated species.

### 5.4 Thermal/non-thermal Desorption

The observed D/H ratios could have been set during an earlier, colder phase of the proto-BD's history and preserved on the dust grains. Once the core gets warmer than ~20 K, CO sublimation and the endothermic exchange reaction $H_2D^+ + H_2 \rightarrow H_3^+ + HD$ becomes effective in the destruction of $H_3^+$ and terminating the extreme deuteration, thus reducing the D/H ratios. As the temperature increases, depletions of heavy species onto grains are less and less likely. Thermal and/or non-thermal desorption becomes important and can, for a limited time, enhance deuteration in the gas phase as the results of previous fractionation on grain surfaces transfer to the gas.

The typical desorption mechanisms are thermal desorption, desorption due to cosmic-ray heating (CRH) of grains, photo-desorption, and chemi-desorption. Chemi-desorption can further populate the gas-phase of species newly formed on grain surfaces. The exact amount is however still debated (e.g., Minissale et al. 2016). We note that no X-ray emission has been detected yet in proto-BDs, due to which non-thermal desorption due to CRH is not expected to be enhanced in these objects but standard CRH must be present. Since the temperature throughout a Class 0/I proto-BD system is expected to be ≤10 K, too low for efficient evaporation of ice mantles, these deuterated molecules are likely to have been returned to the gas phase later during the protostellar stage via some form of non-thermal desorption mechanism, rather than thermal desorption.

The molecules released due to non-thermal desorption are processed in the gas and could even be re-accreted onto the grains. The possible effects of chemical processing and radial location of the origin of emission cannot be disentangled when measuring the D/H ratios derived from single-dish observations. Another efficient mechanism of non-thermal desorption is the presence of a hard shock front very close to the driving source making the close environs warmer than the rest. J163136 is a good test case for this mechanism. The near-infrared spectro-images of the outflow driven by this proto-BD show a very bright shock close (<0.5″; < 70 au) to the driving source (Riaz & Bally 2021). This is also the only proto-BD in our sample with a non-detection in the $DCO^+$ line. The bright shock front could have warmed up the environment close to the source, resulting in enhanced photo-desorption. Future dedicated studies of chemical models for proto-BDs with a deuterated network and including the effects of the various desorption mechanisms can provide a better interpretation of our results.

### 5.5 Evolutionary Timescales

The high D/H ratios for the proto-BDs suggest that they form in cold environments with longer evolutionary timescales and thus longer free-fall timescales through their pre-(sub)stellar stages, due to which gas and dust spend longer times at low temperatures. The D/H ratios are higher in case of stronger depletion of heavy molecules and lower degree of ionization, and are also expected to be higher for more chemically evolved cores (e.g., Aikawa et al. 2012). As shown in the models by e.g., Turner (2001) and Aikawa et al. (2001; 2003), there are three competing time-dependent processes: gas-phase deuterium fractionation, depletion of molecules onto dust grain surface, and dynamical timescale. The chemical timescale is comparable to or longer than the dynamical timescale of the clouds. Other than the noted differences in the origin of emission, the lack of correlation between the DNC/HNC, DNC/$HN^{13}C$, $N_2D^+$/$N_2H^+$ ratios and the CO depletion factor for the proto-BDs could be due to differences in the timescale for deuterium fractionation and the depletion timescale (~$10^5$ yr; Turner 2001). It may be the case that deuterium fractionation does not follow the change in depletion in a dynamically evolving proto-BD core, or some species may have a faster timescale for the deuterium fractionation than others.

Simulations on brown dwarf formation suggest a much shorter embedded (Class 0/I) lifetime of <0.01 Myr for proto-BDs compared to the typical ~0.1 Myr embedded lifetime estimated for low-mass protostars (e.g., Evans et al. 2009; Machida et al. 2009). Indeed, based on physical+chemical modelling of ALMA molecular line observations, the kinematical age is estimated to be ~6000 yr for a Class 0 proto-BD, and ~30,000 yr for a Class I proto-BD (Riaz et al. 2019; 2021). The fact that similar deuteration levels are found in Class 0/Stage 0 and Class I/Stage I proto-BDs, with no notable difference in the detection/non-detection or molecular abundances of these deuterated species, suggests a rapid transition from the Class 0 to Class I stage in sub-stellar objects (e.g., Riaz

& Machida 2021). The high D/H ratios we have measured indicate a rapid deuterium fractionation from the ISM value right after the formation of the proto-BDs, while the high ratio in a Class II brown dwarf (J041858) suggests that the enhanced D/H ratios are present past the early formation stages.

## 6 SUMMARY

We have conducted the first extensive observational survey of several deuterated species in 16 Class 0/I proto-BDs and 4 Class Flat/Class II brown dwarfs. Among the Class 0/I proto-BDs, 15 show emission in $DCO^+$ (3-2), 7 in DCN (3-2), 9 in DNC (3-2), and 5 in $N_2D^+$ (3-2) line. Among the Class Flat/Class II brown dwarfs, only one object shows emission in the $DCO^+$ (3-2) line and weak emission in the DNC (3-2) line.

• The mean D/H ratios for the proto-BDs derived from the deuterated and non-deuterated isotopologues of these molecules are: $DCO^+/HCO^+$ = 0.18±0.19, DCN/HCN = 0.16±0.14, DNC/HNC = 0.19±0.25, and $N_2D^+/N_2H^+$ = 0.02±0.008.

• The $DCO^+/H^{13}CO^+$ (~0.3–4.7), $DCN/H^{13}CN$ (0.4-0.6), and $DNC/HN^{13}C$ (1.2-6.6) ratios are comparatively higher and show a narrower range than the $DCO^+/HCO^+$ (~0.008–0.6), DCN/HCN (0.02-0.4), and DNC/HNC (0.02-0.8) ratios, respectively.

• No correlation is seen between the various D/H ratios and the CO depletion factor. The only exception are the DCN/HCN ratios that show a relatively strong anti-correlation with the CO depletion factor, which can be explained by the difference in the peak emitting regions of the DCN and HCN molecules.

• No correlation is seen between the various D/H ratios and the evolutionary stage of the system.

• Among the proto-BDs, there is a clear rise in $DCO^+/HCO^+$ but a decline in the $N_2D^+/N_2H^+$ ratios with decreasing $L_{bol}$, whereas there is no correlation between the DCN/HCN and DNC/HNC ratios and $L_{bol}$.

• Over a wide range in the bolometric luminosities spanning ~0.002–40 $L_\odot$, we find a trend of higher $DCO^+/HCO^+$ (r = -0.7) and DCN/HCN (r = -0.6) ratios, nearly constant DNC/HNC (r = -0.4) and $DNC/HN^{13}C$ (r = -0.1) ratios, lower $N_2D^+/N_2H^+$ ratios (r = 0.6) in the proto-BDs compared to protostars.

• The D/H ratios for the proto-BDs show the most overlap with the measurements in low-mass protostars and cold molecular clouds. The high D/H ratios for the proto-BDs suggest that both low-temperature <20 K gas-phase ion-molecule deuteron transfer reactions and grain surface reactions can result in enhanced deuterium fractionation in these objects. The very dense and cold ($n_{H_2}$ $\geq 10^6$ cm$^{-3}$, T $\leq$10 K) interior of the proto-BDs provide the ripe conditions for efficient deuterium fractionation in these cores.

• The $DCO^+/HCO^+$, $N_2D^+/N_2H^+$, DCN/HCN and DNC/HNC ratios for the proto-BDs are within the range predicted by current chemical models for low-mass protostars. Future work on chemical models with conditions specific to proto-BDs are required for a more robust comparison.

## ACKNOWLEDGEMENTS

We thank the referee, Jennifer Hatchell, for a detailed review and suggestions on the paper. We also thank Adam Burrows for an insightful discussion on the D/H ratios in the brown dwarf evolutionary models. B.R. acknowledges funding from the Deutsche Forschungsgemeinschaft (DFG) - Projekt number RI-2919/2-1. This work is based on observations carried out with the IRAM 30m telescope. IRAM is supported by INSU/CNRS (France), MPG (Germany) and IGN (Spain).

## 7 DATA AVAILABILITY

The data underlying this article are available in the IRAM archives through the VizieR online database.

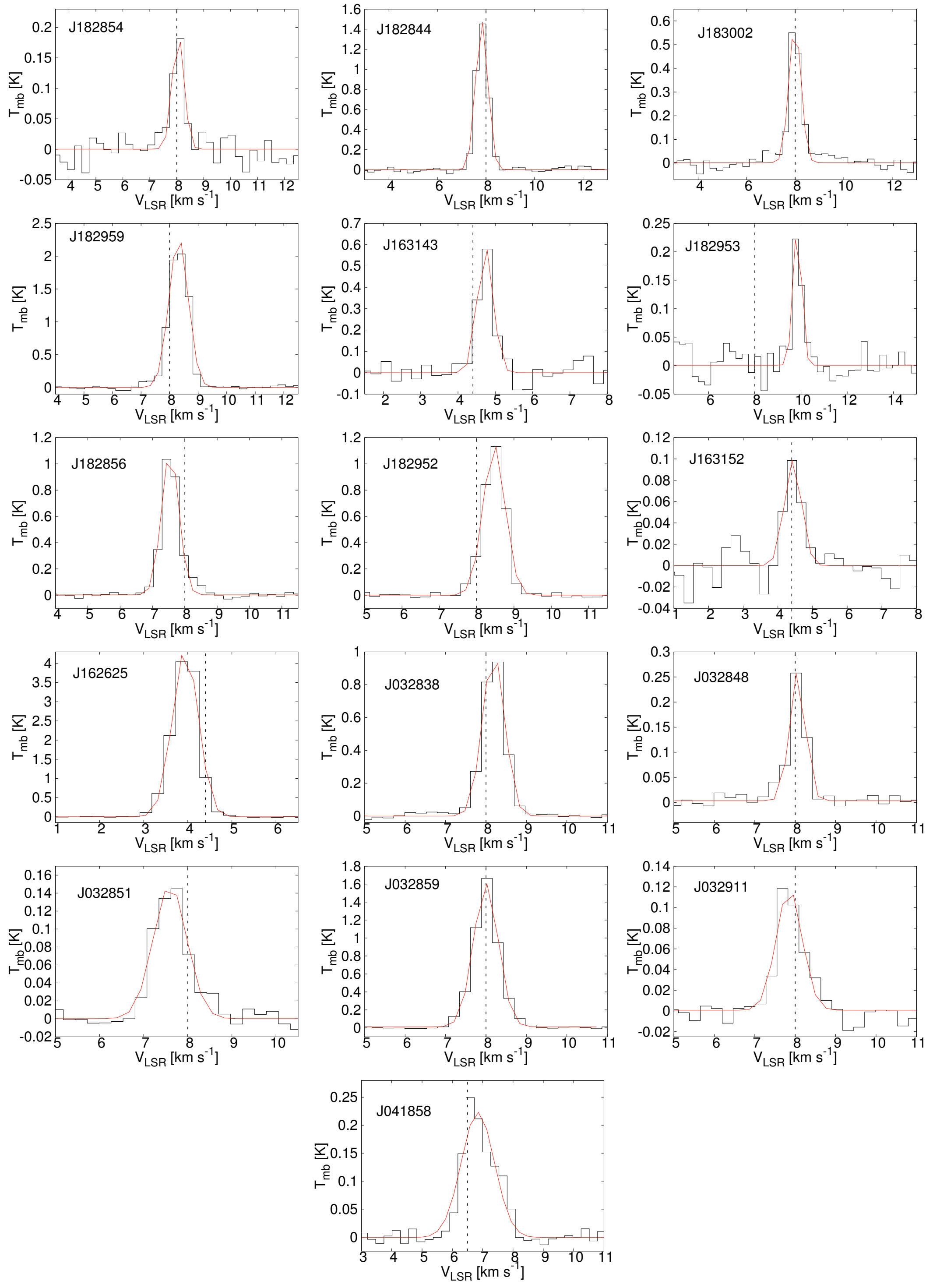

**Figure A1.** The observed $DCO^+$ (3-2) spectra (black) with the Gaussian fits (red). Black dashed line marks the cloud systemic velocity of ~8 km/s in Serpens and Perseus, ~4.4 km/s in Ophiuchus, and ~6.5 km/s in Taurus.

## APPENDIX A: SPECTRA AND LINE PARAMETERS

The observed $DCO^+$ (3-2), DCN (3-2), DNC (3-2), $N_2D^+$ (3-2), $N_2H^+$ (3-2) spectra are shown in Figs. A1, A2, A3, A4, A5, respectively. The line parameters derived from these spectra are listed in Tables A1, A2, A3, A4, A5.

This paper has been typeset from a TEX/LATEX file prepared by the author.

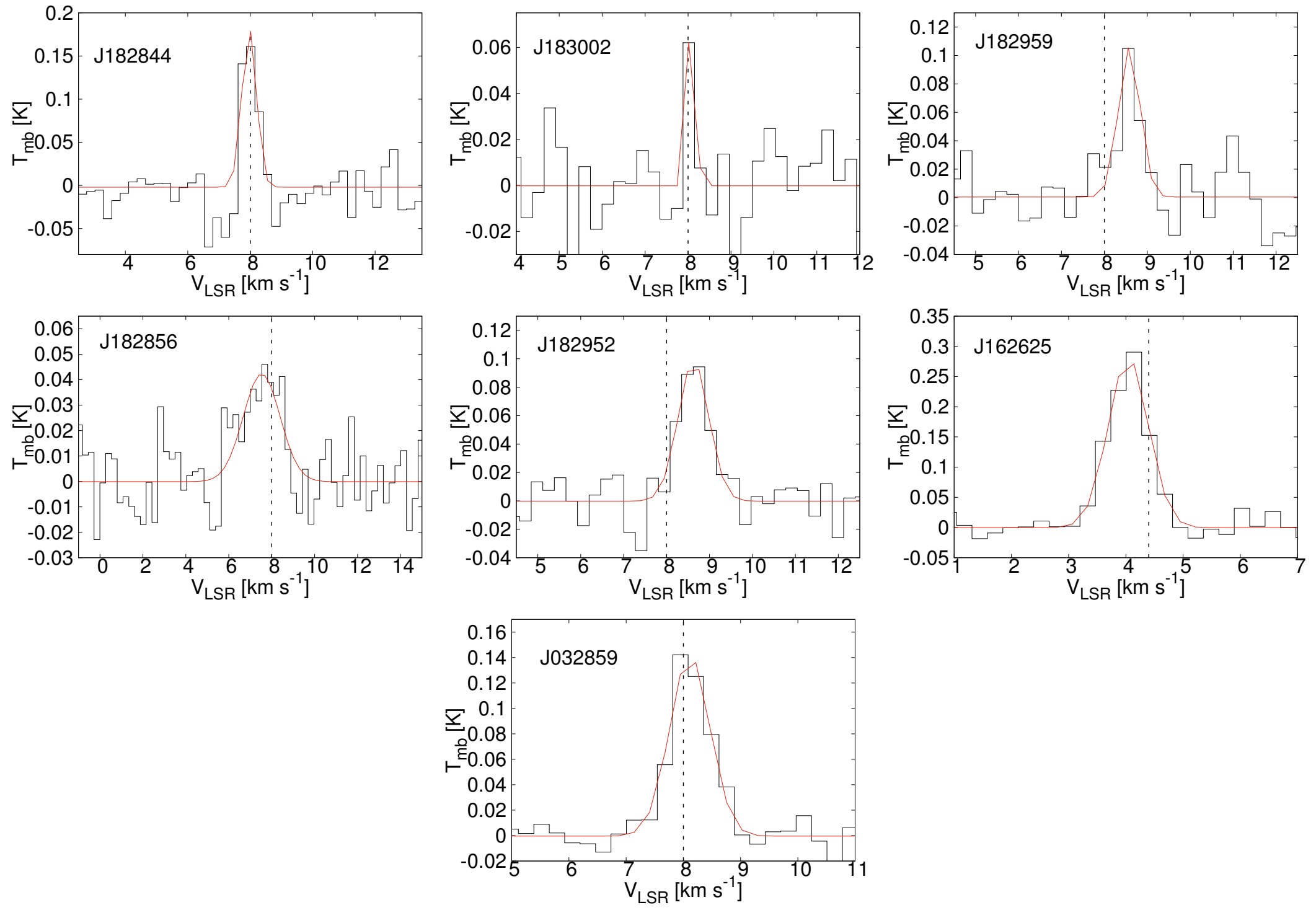

**Figure A2.** The observed DCN (3-2) spectra (black) with the Gaussian fits (red). Black dashed line marks the cloud systemic velocity of ~8 km/s in Serpens and Perseus, ~4.4 km/s in Ophiuchus, and ~6.5 km/s in Taurus.

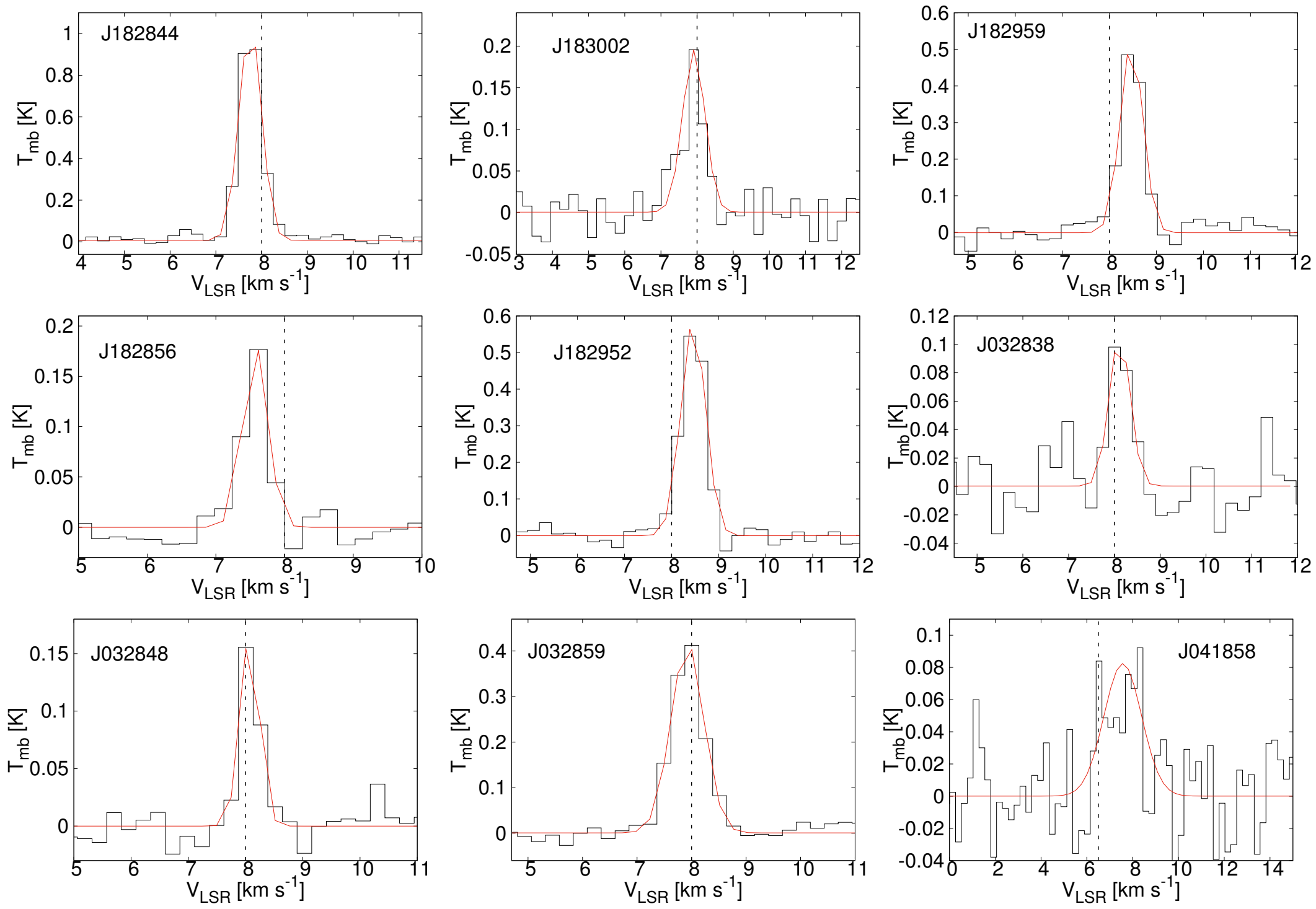

**Figure A3.** The observed DNC (3-2) spectra (black) with the Gaussian fits (red). Black dashed line marks the cloud systemic velocity of ~8 km/s in Serpens and Perseus, ~4.4 km/s in Ophiuchus, and ~6.5 km/s in Taurus. J041858 shows weak emission at a ~2-σ level in this line.

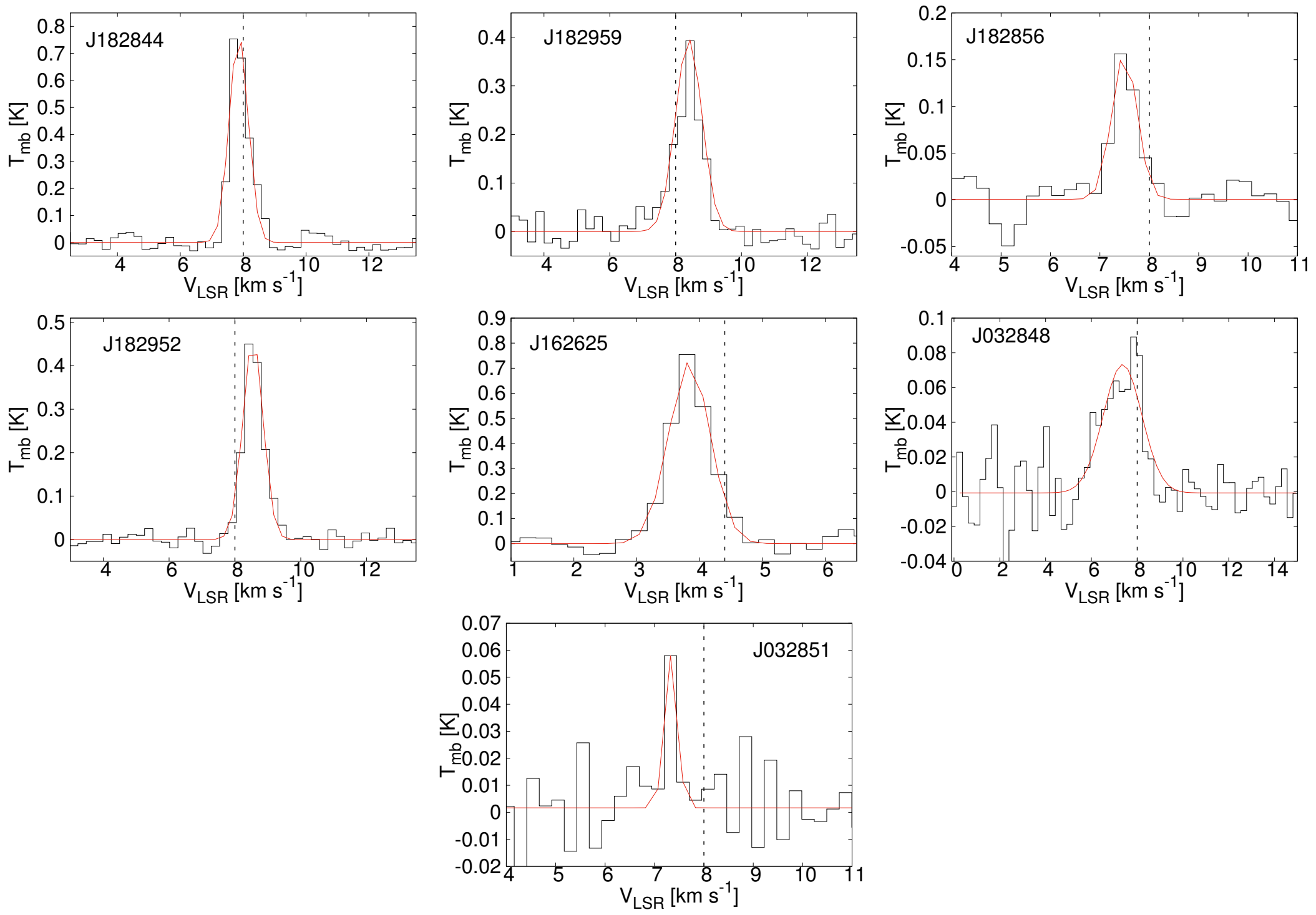

**Figure A4.** The observed $N_2D^+$ (3-2) spectra (black) with the Gaussian fits (red). Black dashed line marks the cloud systemic velocity of ~8 km/s in Serpens and Perseus, ~4.4 km/s in Ophiuchus, and ~6.5 km/s in Taurus.

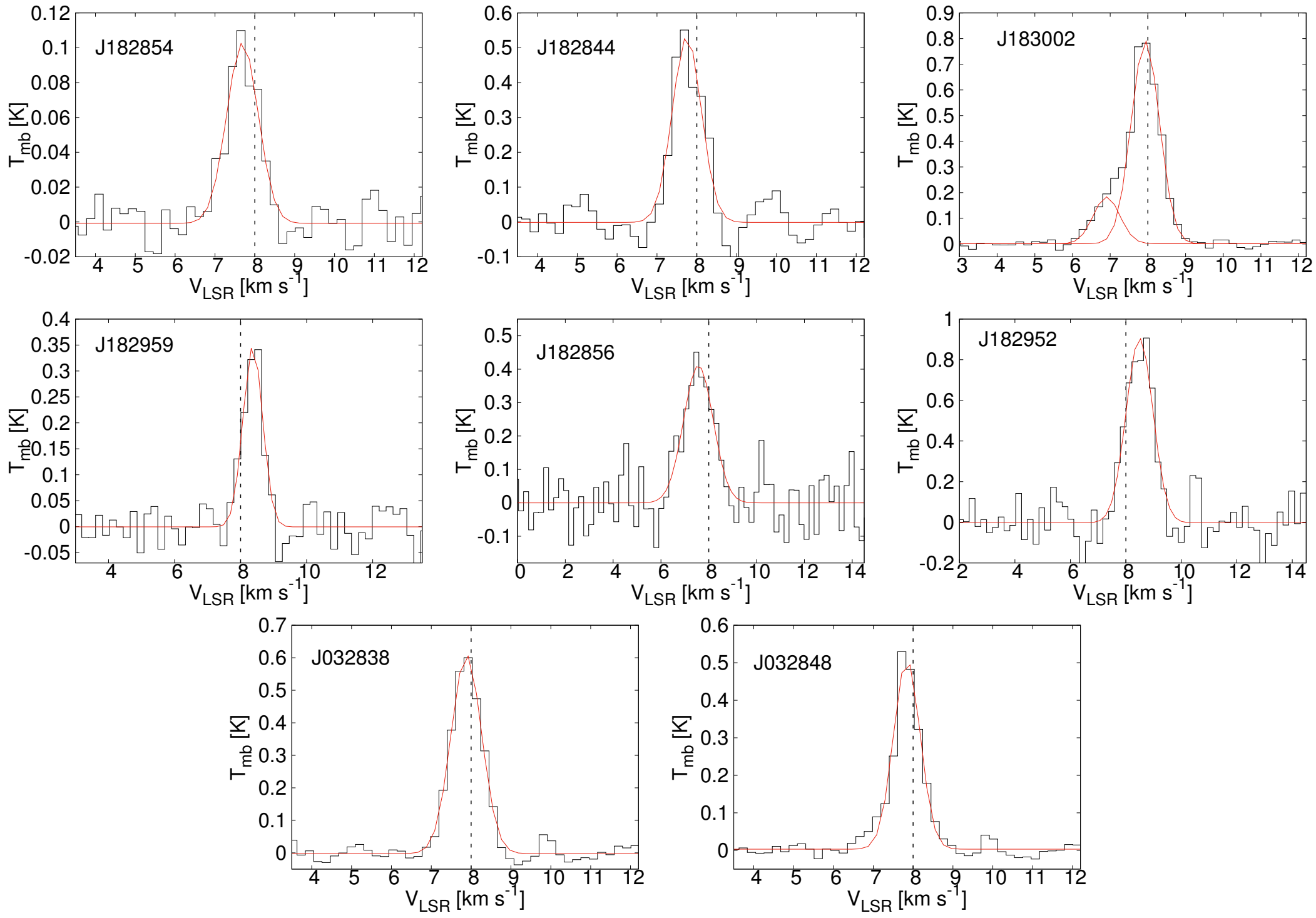

**Figure A5.** The observed $N_2H^+$ (3-2) spectra (black) with the Gaussian fits (red). Black dashed line marks the cloud systemic velocity of ~8 km/s in Serpens and Perseus, ~4.4 km/s in Ophiuchus, and ~6.5 km/s in Taurus.

**Table A1.** $DCO^+$ (3-2) Line Parameters

| Object | $V_{lsr}$ (km $s^{-1}$) | $T_{mb}$ (K) | $\int T_{mb}dv$ (K km $s^{-1}$) | $\Delta v$ (km $s^{-1}$) |
|---|---|---|---|---|
| J182854 | 8.0 | 0.2 | 0.1 | 0.4 |
| J182844 | 7.8 | 1.5 | 0.9 | 0.5 |
| J183002 | 7.9 | 0.5 | 0.4 | 0.5 |
| J182959 | 8.3 | 2.1 | 1.8 | 0.6 |
| J163143 | 4.7 | 0.6 | 0.3 | 0.4 |
| J182953 | 9.8 | 0.2 | 0.1 | 0.4 |
| J163136 | 4.4 | <0.04 | <0.04 | 1.0 |
| J182856 | 7.4 | 1.03 | 0.7 | 0.7 |
| J182952 | 8.5 | 1.1 | 0.8 | 0.7 |
| J163152 | 4.4 | 0.09 | 0.06 | 0.6 |
| J162625 | 3.8 | 4.0 | 3.2 | 0.8 |
| J032838 | 8.3 | 0.9 | 0.7 | 0.7 |
| J032848 | 8.0 | 0.2 | 0.2 | 0.6 |
| J032851 | 7.7 | 0.1 | 0.1 | 0.9 |
| J032859 | 8.0 | 1.6 | 1.3 | 0.8 |
| J032911 | 7.7 | 0.1 | 0.1 | 0.8 |
| J041858 | 6.6 | 0.2 | 0.3 | 1.2 |
| J182940 | 8.0 | <0.03 | <0.03 | 1.0 |
| J182927 | 8.0 | <0.05 | <0.05 | 1.0 |
| J182952 | 8.0 | <0.04 | <0.04 | 1.0 |

The uncertainty is estimated to be ~15%-20% for the integrated intensity, ~6%-8% on $V_{lsr}$, ~5%-10% on $T_{mb}$ and $\Delta v$.

**Table A2.** DCN (3-2) Line Parameters

| Object | $V_{lsr}$ (km $s^{-1}$) | $T_{mb}$ (K) | $\int T_{mb}dv$ (K km $s^{-1}$) | $\Delta v$ (km $s^{-1}$) |
|---|---|---|---|---|
| J182854 | 8.0 | <0.05 | <0.05 | 1.0 |
| J182844 | 7.9 | 0.2 | 0.05 | 0.5 |
| J183002 | 8.1 | 0.07 | 0.02 | 0.1 |
| J182959 | 8.6 | 0.1 | 0.07 | 0.5 |
| J163143 | 8.0 | <0.03 | <0.03 | 1.0 |
| J182953 | 8.0 | <0.05 | <0.05 | 1.0 |
| J163136 | 8.0 | <0.04 | <0.04 | 1.0 |
| J182856 | 7.7 | 0.05 | 0.09 | 1.6 |
| J182952 | 8.7 | 0.09 | 0.08 | 0.8 |
| J163152 | 4.4 | <0.02 | <0.02 | 1.0 |
| J162625 | 4.1 | 0.3 | 0.2 | 0.8 |
| J032838 | 8.0 | <0.03 | <0.03 | 1.0 |
| J032848 | 8.0 | <0.03 | <0.02 | 1.0 |
| J032859 | 7.9 | 0.1 | 0.1 | 0.9 |
| J032911 | 8.0 | <0.02 | <0.02 | 1.0 |
| J041858 | 6.5 | <0.01 | <0.01 | 1.0 |
| J182940 | 8.0 | <0.03 | <0.03 | 1.0 |
| J182927 | 8.0 | <0.05 | <0.05 | 1.0 |
| J182952 | 8.0 | <0.04 | <0.04 | 1.0 |

The uncertainty is estimated to be ~15%-20% for the integrated intensity, ~6%-8% on $V_{lsr}$, ~5%-10% on $T_{mb}$ and $\Delta v$.

**Table A3.** DNC (3-2) Line Parameters

| Object | $V_{lsr}$ (km $s^{-1}$) | $T_{mb}$ (K) | $\int T_{mb}dv$ (K km $s^{-1}$) | $\Delta v$ (km $s^{-1}$) |
|---|---|---|---|---|
| J182854 | 8.0 | <0.05 | <0.05 | 1.0 |
| J182844 | 7.8 | 0.91 | 0.7 | 0.6 |
| J183002 | 7.9 | 0.1 | 0.07 | 0.4 |
| J182959 | 8.4 | 0.5 | 0.3 | 0.5 |
| J163143 | 8.0 | <0.03 | <0.03 | 1.0 |
| J182953 | 8.0 | <0.05 | <0.05 | 1.0 |
| J163136 | 8.0 | <0.04 | <0.04 | 1.0 |
| J182856 | 7.6 | 0.2 | 0.08 | 0.5 |
| J182952 | 8.4 | 0.5 | 0.4 | 0.6 |
| J163152 | 4.4 | <0.02 | <0.02 | 1.0 |
| J032838 | 8.0 | 0.1 | 0.06 | 0.6 |
| J032848 | 8.0 | 0.1 | 0.07 | 0.5 |
| J032851 | 8.0 | <0.03 | <0.03 | 1.0 |
| J032859 | 8.0 | 0.4 | 0.3 | 0.8 |
| J032911 | 8.0 | <0.06 | <0.06 | 1.0 |
| J041858 | 7.6 | <0.08 | <0.13 | 1.6 |
| J182940 | 8.0 | <0.03 | <0.03 | 1.0 |
| J182927 | 8.0 | <0.05 | <0.05 | 1.0 |
| J182952 | 8.0 | <0.04 | <0.04 | 1.0 |

The uncertainty is estimated to be ~15%-20% for the integrated intensity, ~6%-8% on $V_{lsr}$, ~5%-10% on $T_{mb}$ and $\Delta v$.

**Table A4.** $N_2D^+$ (3-2) Line Parameters

| Object | $V_{lsr}$ (km $s^{-1}$) | $T_{mb}$ (K) | $\int T_{mb}dv$ (K km $s^{-1}$) | $\Delta v$ (km $s^{-1}$) |
|---|---|---|---|---|
| J182854 | 8.0 | <0.05 | <0.05 | 1.0 |
| J182844 | 7.8 | 0.7 | 0.6 | 0.6 |
| J183002 | 8.0 | <0.05 | <0.05 | 1.0 |
| J182959 | 8.3 | 0.3 | 0.3 | 0.8 |
| J163143 | 8.0 | <0.03 | <0.03 | 1.0 |
| J182953 | 8.0 | <0.05 | <0.05 | 1.0 |
| J163136 | 8.0 | <0.04 | <0.04 | 1.0 |
| J182856 | 7.4 | 0.2 | 0.1 | 0.6 |
| J182952 | 8.4 | 0.4 | 0.4 | 0.8 |
| J163152 | 4.4 | <0.02 | <0.02 | 1.0 |
| J162625 | 3.8 | 0.7 | 0.6 | 0.8 |
| J032838 | 8.0 | <0.06 | <0.06 | 1.0 |
| J032848 | 7.8 | 0.1 | 0.1 | 1.7 |
| J032851 | 7.3 | 0.06 | 0.04 | 0.6 |
| J032911 | 8.0 | <0.03 | <0.03 | 1.0 |
| J041858 | 6.5 | <0.02 | <0.02 | 1.0 |
| J182940 | 8.0 | <0.03 | <0.03 | 1.0 |
| J182927 | 8.0 | <0.05 | <0.05 | 1.0 |
| J182952 | 8.0 | <0.04 | <0.04 | 1.0 |

The uncertainty is estimated to be ~15%-20% for the integrated intensity, ~6%-8% on $V_{lsr}$, ~5%-10% on $T_{mb}$ and $\Delta v$.

**Table A5.** $N_2H^+$ (3-2) Line Parameters

| Object | $V_{lsr}$ (km $s^{-1}$) | $T_{mb}$ (K) | $\int T_{mb}dv$ (K km $s^{-1}$) | $\Delta v$ (km $s^{-1}$) |
|---|---|---|---|---|
| J182854 | 7.6 | 0.1 | 0.09 | 0.8 |
| J182844 | 7.7 | 0.5 | 0.5 | 0.8 |
| J183002 (peak 1) | 6.9 | 0.2 | 0.2 | 0.7 |
| (peak 2) | 7.9 | 0.8 | 0.7 | 0.7 |
| J182959 | 8.5 | 0.3 | 0.2 | 0.7 |
| J182856 | 7.5 | 0.4 | 0.6 | 1.4 |
| J182952 | 8.7 | 0.9 | 1.0 | 1.1 |
| J032838 | 7.8 | 0.6 | 0.4 | 0.9 |
| J032848 | 7.8 | 0.4 | 0.3 | 0.7 |

The uncertainty is estimated to be ~15%-20% for the integrated intensity, ~6%-8% on $V_{lsr}$, ~5%-10% on $T_{mb}$ and $\Delta v$.

**Table A6.** Excitation temperature and optical depth

| Object | $DCO^+$ (3-2) | | DCN (3-2) | | DNC (3-2) | | $N_2D^+$ (3-2) | | $N_2H^+$ (3-2) | |
|---|---|---|---|---|---|---|---|---|---|---|
| | $T_{ex}$ (K) | $\tau$ | $T_{ex}$ (K) | $\tau$ | $T_{ex}$ (K) | $\tau$ | $T_{ex}$ (K) | $\tau$ | $T_{ex}$ (K) | $\tau$ |
| J182854 | 7.0 | 0.1 | – | – | – | – | – | – | 6.0 | 0.1 |
| J182844 | 7.0 | 0.7 | 4.7 | 0.2 | 5.0 | 1.4 | 7.0 | 0.3 | 5.8 | 0.5 |
| J183002 | 8.3 | 0.1 | 6.0 | 0.04 | 6.0 | 0.1 | – | – | 7.0 | 0.4 |
| J182959 | 7.4 | 1.0 | 4.8 | 0.1 | 4.8 | 0.6 | 7.0 | 0.1 | 5.7 | 0.2 |
| J163143 | 5.5 | 0.4 | – | – | – | – | – | – | – | – |
| J182953 | 6.0 | 0.1 | – | – | – | – | – | – | – | – |
| J163136 | – | – | – | – | – | – | – | – | – | – |
| J182856 | 7.8 | 0.3 | 5.4 | 0.04 | 8.6 | 0.04 | 9.7 | 0.03 | 6.5 | 0.3 |
| J182952 | 9.0 | 0.3 | 6.6 | 0.03 | 9.5 | 0.1 | 9.8 | 0.09 | 8.0 | 0.4 |
| J163152 | 7.7 | 0.03 | – | – | – | – | – | – | – | – |
| J162625 | 9.5 | 2.1 | 6.6 | 0.1 | – | – | 9.8 | 0.1 | – | – |
| J032838 | 9.3 | 0.2 | – | – | 8.0 | 0.03 | – | – | 8.8 | 0.2 |
| J032848 | 9.0 | 0.07 | – | – | 7.0 | 0.06 | 9.0 | 0.02 | 8.0 | 0.2 |
| J032851 | 8.7 | 0.03 | – | – | – | – | 8.7 | 0.01 | – | – |
| J032859 | 9.7 | 0.4 | 8.6 | 0.03 | 8.7 | 0.1 | – | – | – | – |
| J032911 | 9.2 | 0.02 | – | – | – | – | – | – | – | – |
| J041858 | 7.5 | 0.08 | – | – | 5.2 | 0.04 | – | – | – | – |

The uncertainty is estimated to be ~20% on $T_{ex}$ and $\tau$ estimates.